\documentclass[sigconf, nonacm]{acmart}

\newcommand\vldbdoi{10.14778/3446095.3446100}

\newcommand\vldbvolume{14}
\newcommand\vldbissue{5}
\newcommand\vldbyear{2021}
\newcommand\vldbauthors{\authors}
\newcommand\vldbtitle{\shorttitle} 
\newcommand\vldbavailabilityurl{https://github.com/msr-fiddle/DS-Analyzer}
\newcommand\vldbpagestyle{empty}

\usepackage{tikz}
\usepackage{amsmath}
\usepackage{adjustbox}
\usepackage{amsthm}
\usepackage{caption}
\usepackage{changepage}
\usepackage{datetime}
\usepackage{enumitem}
\usepackage{graphicx}
\usepackage{framed}
\usepackage{balance}
\usepackage{multirow}
\usepackage{url}

\usepackage{breakurl}
\usepackage{verbatim} 
\usepackage{xspace} 
\usepackage[normalem]{ulem}
\usepackage{booktabs,array,ragged2e}
\usepackage[font=large, position=bottom]{subfig}
\usepackage{comment}

\usepackage{color, colortbl}
\definecolor{grey}{gray}{0.9}
\definecolor{lgrey}{gray}{1}
\definecolor{white}{rgb}{1,1,1}
\definecolor{dgreen}{HTML}{228B22}

\newlength\mylength
\newcommand{\mycaption}[2]{\caption{\textbf{#1}. {#2}}}
\newcommand{\sref}[1]{\S\ref{#1}}
\newcommand{\vheading}[1]{\vspace{0.05in}\noindent\textbf{#1}}
\newcommand{\myx}{$\times$\xspace}
\newcommand{\vtt}[1]{\texttt{#1}\xspace}

\newcommand{\dl}{{DataLoader}\xspace}

\newcommand{\minio}{MinIO\xspace}

\newcommand{\sysname}{CoorDL\xspace}
\newcommand{\pysysname}{Py-CoorDL\xspace}

\newcommand{\eg}{\textit{e.g.,}\xspace}
\newcommand{\etal}{\textit{et al.}\xspace}

\newcommand{\gb}{GiB\xspace}

\newcommand{\tb}{TiB\xspace}
\newcommand{\confssd}{\texttt{Config-SSD-V100}\xspace}
\newcommand{\confhdd}{\texttt{Config-HDD-1080Ti}\xspace}
\newcommand{\tool}{{DS-Analyzer}\xspace}
\newcommand{\begincompactitemize}{\begin{itemize}[noitemsep,topsep=0pt,parsep=0pt,partopsep=0pt]}
\newcommand{\vio}{{I/O}\xspace}

\usepackage{hyperref}

\usepackage{ifthen}
\newboolean{publicversion}
\setboolean{publicversion}{true}

\newboolean{addappendix}
\setboolean{addappendix}{true}

\IfFileExists{intro-flag}
{
	\setboolean{publicversion}{True}
}
{}

\ifthenelse{\boolean{publicversion}}{
	\newcommand{\grumbler}[3]{}
}{
	\newcommand{\grumbler}[3]{\xspace\textcolor{#3}{\bf #1: #2}}
}

\newcommand{\ra}[1]{\renewcommand{\arraystretch}{#1}}

\newcommand{\vc}[1]{\grumbler{Vijay}{#1}{red}}
\newcommand{\jm}[1]{\grumbler{Jayashree}{#1}{magenta}}

\newcommand{\revision}[1]{\xspace{#1}}

\makeatletter
\def\balanceissued{unbalanced}
\let\oldbibitem\bibitem
\def\bibitem{%
	\ifnum\thepage=\theTotPages{}%
	\expandafter\ifx\expandafter\relax\balanceissued\relax\else%
	\balance%
	\gdef\balanceissued{\relax}\fi%
	\else\fi%
	\oldbibitem}
\makeatother

\begin{document}
	
	\title{Analyzing and Mitigating Data Stalls in DNN Training}
	
	\newcommand*\samethanks[1][\value{footnote}]{\footnotemark[#1]}
	\newcommand{\ack}{{\large$^\ast$}}
	\author{Jayashree Mohan*}
	\affiliation{%
		\institution{University of Texas at Austin}
	}
	\thanks{*Work done as part of MSR internship}
	\email{jaya@cs.utexas.edu}
	
	\author{Amar Phanishayee}
	\affiliation{%
		\institution{Microsoft Research}
	}
	\email{amar@microsoft.com}
	
	\author{Ashish Raniwala}
	\affiliation{%
		\institution{Microsoft}
	}
	\email{ashish.raniwala@microsoft.com}
	
	\author{Vijay Chidambaram}
	\affiliation{%
		\institution{University of Texas at Austin \& VMWare Research}
	}
	\email{vijay@cs.utexas.edu}

	
	\begin{abstract}
  Training Deep Neural Networks (DNNs) is resource-intensive and
  time-consuming. While prior research has explored many different 
  ways of reducing DNN training time, 
  the impact of \emph{input data pipeline}, i.e., fetching raw data items from storage and
  performing data pre-processing in memory, has been relatively unexplored.  
  This paper makes the following contributions: (1) We present the
  first comprehensive analysis of how the input data pipeline affects
  the training time of widely-used \revision{computer vision and audio} 
  Deep Neural Networks (DNNs), \revision{that typically involve complex
  data pre-processing}. We
  analyze nine different models across three tasks 
  and four datasets
  while varying factors such as the amount of memory, number of CPU
  threads, storage device, GPU generation etc on servers that are a
  part of a large production cluster at Microsoft. We find that in
  many cases, DNN training time is dominated by \emph{data stall
    time}: time spent waiting for data to be fetched and
  pre-processed. (2) We build a tool, \tool to precisely measure data
  stalls using a differential technique, and perform predictive
  what-if analysis on data stalls.  (3) Finally, based on the insights
  from our analysis, we design and implement three simple but
  effective techniques in a data-loading library, \sysname, to
  mitigate data stalls.  Our experiments on a range of DNN tasks,
  models, datasets, and hardware configs show that when PyTorch
  uses \sysname instead of the state-of-the-art DALI data loading
  library, DNN training time is reduced significantly (by as much as
  5\myx on a single server). 
  

\end{abstract}
	\maketitle

		\pagestyle{\vldbpagestyle}
		\begingroup\small\noindent\raggedright\textbf{PVLDB Reference Format:}\\
		\vldbauthors. \vldbtitle. PVLDB, \vldbvolume(\vldbissue): 
		\vldbyear.\\
		\href{https://doi.org/\vldbdoi}{doi:\vldbdoi}
		\endgroup
		\begingroup
		\renewcommand\thefootnote{}\footnote{\noindent
			This work is licensed under the Creative Commons BY-NC-ND 4.0 International License. Visit \url{https://creativecommons.org/licenses/by-nc-nd/4.0/} to view a copy of this license. For any use beyond those covered by this license, obtain permission by emailing \href{mailto:info@vldb.org}{info@vldb.org}. Copyright is held by the owner/author(s). Publication rights licensed to the VLDB Endowment. \\
			\raggedright Proceedings of the VLDB Endowment, Vol. \vldbvolume, No. \vldbissue\ %
			ISSN 2150-8097. \\
			\href{https://doi.org/\vldbdoi}{doi:\vldbdoi} \\
		}\addtocounter{footnote}{-1}\endgroup
		
		\ifdefempty{\vldbavailabilityurl}{}{
			\vspace{.3cm}
			\begingroup\small\noindent\raggedright\textbf{PVLDB Artifact Availability:}\\
			The source code, data, and/or other artifacts have been made available at \url{\vldbavailabilityurl}.
			\endgroup
		}

	\section{Introduction}
\label{sec-intro}

Data is the fuel powering machine learning~\cite{andrew-ng}. 
Large training datasets are empowering state-of-the-art accuracy
for several machine learning tasks. Particularly, Deep Neural Networks (DNNs), have gained
prominence, as they allow us to tackle problems that were previously
intractable, such as image
classification~\cite{krizhevsky2012imagenet,simonyan2014very,he2016deep},
translation~\cite{wu2016google}, speech
recognition\cite{graves2013speech}, video
captioning~\cite{venugopalan2015sequence}, and even predictive
health-care~\cite{deepmind-health}.

Empowering DNNs to push state-of-the-art accuracy requires the model to
be trained with a large volume of data.
 During training, the model predicts the output given training data; based on
the output, the model's weights are tuned. This happens iteratively,
in many rounds called epochs. 

However, DNN training is data-hungry, resource-intensive, and
time-consuming. It involves the holistic use of all
the resources in a server from storage and CPU for 
fetching and pre-processing the dataset to the GPUs that perform 
computation on the transformed data. 
Researchers have tackled how to efficiently use these
resources to reduce DNN training time,
such as reducing communication
overhead~\cite{lin2017dgc, PoseidonATC2017, hashemi2018tictac,
	jayarajan2019priority, narayanan2019pipedream}, GPU memory
optimizations~\cite{vdnn,chen2016training, gist2018}, and
compiler-based operator optimizations~\cite{vasilache2018tensor,
	chen2018tvm,jia2019taso}.
However, the impact of storage systems, specifically the \emph{data
	pipeline}, on DNN training has been relatively unexplored.

\newcommand\VRule[1][\arrayrulewidth]{\vrule width #1}
\setlength{\tabcolsep}{3pt}
\newcolumntype{M}[1]{>{\centering\arraybackslash}m{#1}}
\begin{table*}[!t]
  \small
  \centering
  \ra{1.0}
  \caption{Key findings and implications of our analysis of data stalls}
  \vspace{-1em}
 \begin{tabular}{!{\VRule[1pt]}M{0.43\textwidth}!{\VRule[1pt]}M{0.54\textwidth}!{\VRule[1pt]}}
	\specialrule{1.2pt}{0pt}{0pt}
	\rowcolor{grey} \centering
    \textbf{\emph{Finding}} & \Centering\textbf{\emph{Insights}}\\
    \specialrule{0.8pt}{0pt}{0pt}
    \rowcolor{white}
    OS Page Cache is inefficient for DNN training due to thrashing & DNN-aware caching can eliminate thrashing across epochs \\
        \specialrule{0.5pt}{0pt}{0pt}
    DNNs need anywhere between 3 -- 24 CPU cores \textit{per GPU} for data pre-processing & If hardware is upgraded to overcome workload bottlenecks, it must be done carefully with an eye towards designing balanced server SKUs.  \\
        \specialrule{0.5pt}{0pt}{0pt}
    DNNs spend upto 65\% of the epoch time in data pre-processing, primarily on redundant decoding & Decoded data can be cached (as opposed to caching encoded data), if space amplification due to decoding can be addressed\\
    \specialrule{0.5pt}{0pt}{0pt}
    Lack of coordination among local caches lead to
    redundant \vio in distributed training across servers & To overcome local storage \vio bottlenecks, local in-memory caches of servers allocated to a job can be coordinated to fetch data from distributed in-memory caches \\
    \specialrule{0.5pt}{0pt}{0pt}
    \revision{Hyperparameter search workloads} perform 
    redundant \vio \& prep & \revision{Hyperparameter search} jobs must coordinate data fetch \& prep to mitigate data stalls\\
	\specialrule{1.2pt}{0pt}{0pt}
   \end{tabular}
   \vspace{-0.5em}

\label{tbl-summary}
\end{table*}

\vheading{The DNN Data Pipeline}. 
During DNN training, the data pipeline works as follows. Data items
are first fetched from storage and then \emph{pre-processed} in memory. 
For example, for many important and widely-used
classes of DNNs \revision{in computer vision}, there are
several pre-processing steps: data is first decompressed,
and then random perturbations such as cropping the image or rotating
it are performed to improve the model's
accuracy~\cite{perez2017effectiveness}. Once pre-processed, the data items
are sent to the GPUs for processing. One complete pass over the training 
dataset is termed an epoch; models are iteratively trained for several epochs to 
achieve desired accuracy.

The DNN data pipeline operates in parallel with GPU computation. Ideally, the data pipeline should steadily feed 
pre-processed data items to the GPUs to keep them continuously busy
processing data; we term this GPU-bound. Unfortunately, DNN training
is often \vio-bound, bottlenecked by fetching the data from storage, or
CPU-bound, bottlenecked by applying data pre-processsing in
memory. Collectively, we term these bottlenecks \textbf{data stalls} and
differentiate between \emph{prep stalls} (time spent on data pre-processing)
and \emph{fetch stalls} (time spent on \vio).

\vheading{Novelty over prior work}. Recent work like
Quiver~\cite{kumar2020quiver} demonstrate how to speed up DNN
training by efficiently caching data from remote object storage  on local
storage. In contrast, 
our  work  assumes as baseline the setting where all data is
available on local storage (as most publicly available datasets for popular models fit on local storage~\cite{training-big,
	abu2016youtube, kuznetsova2018open, defferrard2016fma, imagenet-22k, russakovsky2015imagenet}, \sref{sec-fetch-remote}), and shows how to further
optimize the data pipeline (\sref{sec-fetch}, \sref{sec-prep}). This work therefore fundamentally improves performance on top of what Quiver can achieve.  While prior work like
Cerebro~\cite{cerebro}, and DeepIO~\cite{zhu2018entropy} have looked
at optimizing data fetch in distributed training, they do not
systematically analyze data stalls in different training scenarios, or
demonstrate how to accelerate single-server training.


\vspace{-2em}
\subsection{ Contributions}

\vheading{Categorizing, measuring, and analyzing data stalls}.
We present the first comprehensive analysis
of data stalls (categorized as fetch and prep stalls) in DNN training.
We analyze nine popular DNN models from three domains
(image classification, object detection, and audio classification) and
four datasets in a production cluster at Microsoft.
We vary factors such as the storage media, amount of data that can be cached in memory, the number of
CPU threads used to fetch and pre-process data,
and GPU generation.  We then analyze how these factors affect the data
pipeline and DNN training. Our analysis finds that data stalls squander away 
the improved performance of faster GPUs, even on ML optimized servers like the DGX-2~\cite{dgx2}.
Revisiting the insights from Stonebraker \etal~\cite{stonebraker81},
our analysis corroborates that relying on OS abstractions (like Page Cache)
is inefficient for DNN workloads. We also find that the data pipelines in popular
training frameworks like PyTorch and TensorFlow are inefficient in their
use of CPU and memory resources,
despite using state-of-the-art data-loading libraries like
DALI~\cite{noauthor_fast_2019} that reduce prep stalls using
GPU-accelerated data pre-processing. 
Table~\ref{tbl-summary}
summarizes the findings and insights of our analysis.

\vheading{Performing predictive what-if analysis of data stalls}.
Performing an analysis of how the data pipeline impacts DNN training
is challenging since DNN training has a high degree of concurrency; it
is hard to isolate the time taken to perform a single task especially
as data prefetching and pre-processing are pipelined with GPU
computation.  We develop a tool, \tool, that uses differential
analysis between runs (\eg comparing a run where data is completely
cached vs when data needs to be fetched from storage) to accurately
identify data-stall bottlenecks. Using the measured data stalls, \tool
answers what-if questions to help practitioners predict and analyze
data stalls (\eg What would be the impact on data stalls if DRAM
capacity increased by $2\times$?).

\vheading{Mitigating data stalls}. We use the insights from our
analysis to identify opportunities for improvement. We build a new
data-loading library, \sysname, that does not require any changes to
the cluster infrastructure. \sysname uses three main techniques to
mitigate data stalls. First, inspired by the pioneering work of
Stonebraker \etal on database caching~\cite{stonebraker81}, we demonstrate
that relying on the OS page cache is sub-optimal for DNN training.  We
implement \minio, a software cache that is specialized for DNN training.
Second, we describe the \emph{partitioned caching} technique to coordinate
the \minio caches of servers involved in distributed training over
commodity network stack.  Third, we discuss the \emph{coordinated prep}
technique to carefully eliminate redundancy in data prep among
concurrent hyperparameter search jobs in a server. We implement these
techniques as part of the user-space library \sysname, built on top of
the state-of-the-art data pipeline DALI~\cite{noauthor_fast_2019}.  We
evaluate \sysname across different models, datasets, and hardware and
show that it can accelerate training by up-to 5\myx on a single server
by mitigating data stalls over DALI.


	\section{Background}
\label{sec-bkgd}

Deep Neural Networks (DNNs) are a class of ML models that
automatically extract higher level features from the input data.  The
DNN is trained over multiple rounds termed \emph{epochs}. Each epoch
processes all items in the dataset exactly once, and consists of
multiple \emph{iterations}; each iteration processes a random,
disjoint subset of the data termed a \emph{minibatch}. The DNN is
trained until a target accuracy is reached. Training a DNN model to reach a
given accuracy consists of two steps: 
\vspace{-0.5em}
\begin{enumerate}[leftmargin=18pt]
\itemsep0em 
\item \vheading{Hyperparameter (HP) search}. There are many parameters for
the learning algorithm that must be provided before the start of
training.  These hyperparameters (for \eg learning rate, its decay,
dropout, and momentum) influence the speed and quality of
learning.  
During the search process, we start several
training jobs; each job trains the model with different
hyperparameters, on each available GPU (or a distributed job across
several GPUs); progress is checked after a few epochs and the
worst-performing candidates are killed and replaced by new jobs with
different hyperparameters that are chosen
algorithmically~\cite{bergstra2012random, jaderberg2017population,
  golovin2017google, hyperband}.
Tuning
hyperparameters is crucial for generating DNN models that have high
accuracy~\cite{hyper-importance}.

\item \vheading{Training the model to target accuracy}.
The second step is to obtain models
with high accuracy by training it with input data, using the
hyperparameters chosen in the previous step.
\end{enumerate}

\subsection{The DNN ETL Requirements}
\label{sec-etl}
In every epoch of training, the input dataset 
is subjected to a ETL (extract-transform-load) before being processed at the GPU
(or any other accelerator). 
The ETL process in the data pipeline of DNN training imposes several unique data 
ordering constraints to ensure model convergence and achieve state-of-the-art accuracy. 
\vspace{-0.5em}
\begin{itemize}[leftmargin=18pt]
	\item The dataset must be shuffled every epoch to ensure the order in which data 
	items are accessed are random in each epoch 
	\item An epoch must use \emph{all} data items in the dataset \emph{exactly} once 
	\item In every epoch, the data transformations(pre-processing) must be random; the same transformed item should not be used across epochs
\end{itemize}
Several prior work have theoretically and empirically demonstrated that
relaxing these constraints will affect the convergence rate of 
SGD~\cite{convergence, chung2017ubershuffle, localsgd, perez2017effectiveness}.
Therefore, in this work, all our experiments abide by the aforementioned ETL 
requirements.

\subsection{DALI : Fast Data Pipelining} State-of-the-art data loading
and pre-processing libraries like DALI can be used as a drop in
replacement for the default dataloaders in frameworks like PyTorch,
TensorFlow, or MxNet. DALI accelerates data pre-processing
operations 
using GPU-accelerated data pre-processing operations. DALI
also \emph{prefetches and pipelines} the data fetch and pre-processing with
the GPU compute, similar to the default dataloader in PyTorch. We empirically
verified that DALI outperforms the default data pipelines of PyTorch, TensorFlow, 
and MxNet. Therefore, throughout this work, unless and otherwise stated,
we use DALI, as it is the strongest baseline.

\section{Data Stalls in DNN Training}
\label{sec-data-stalls}
We now discuss our formulation of data stalls.
Consider the training process of a typical DNN. It
executes the following steps in each iteration of an epoch:
\begin{enumerate}[leftmargin=18pt]
	\itemsep0em 
	\item A minibatch of data items is fetched from storage.
	\item The data items are pre-processed, for \eg, for image classification, data items are decompressed, and then randomly cropped, resized, and flipped.
	\item The minibatch is then processed at the GPU to obtain the
	model's prediction 
	\item A loss function is used to determine how much the prediction
	deviates from the right answer
	\item Model weights are updated using computed gradients
	
\end{enumerate}

\begin{figure}[!t]
  \centering 
     \includegraphics[width=.48\textwidth]{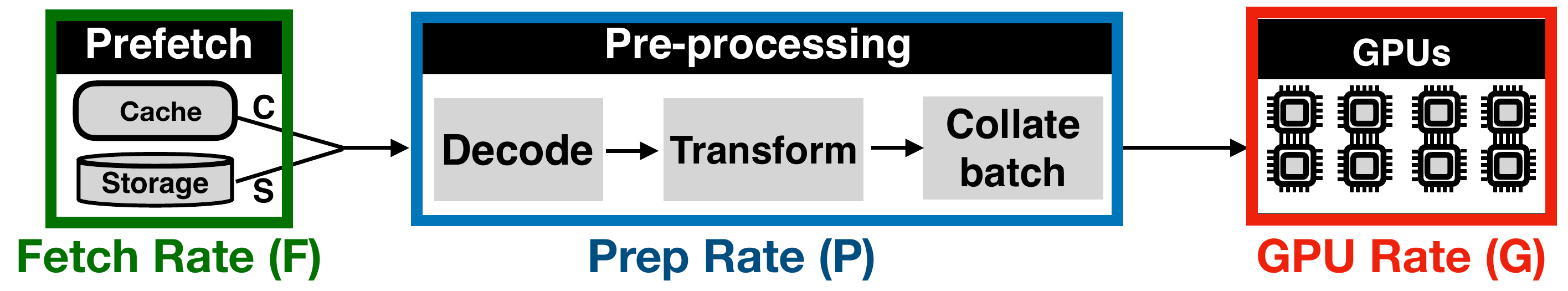}
  \vspace{-2.3em}
  \mycaption{Data Pipeline in DNN training}{This figure shows the different stages in the data pipeline.}
  \vspace{-1.5em}
  \label{fig-rate}

\end{figure}

Ideally, most of the time in each epoch should be spent on Steps 3--5
(which we collectively term the \textit{GPU compute} time), i.e.,
training is \emph{GPU bound}. When performing multi-GPU training,
individual GPUs (workers) exchange weight gradients with other workers
before performing weight update.  For this work, we
roll the communication time for gradient exchange during multi-GPU
training into computation time. 

In most frameworks including PyTorch, TensorFlow, and MxNet,
data preparation (Steps 1 and 2) and GPU computation execute in a
pipelined fashion; i.e., subsequent minibatches are \emph{prefetched and
	pre-processed} by data preparation threads, using multiple CPU cores on
the machine, as the GPU computes on the current minibatch of data.  If
the GPU is waiting for Steps 1--2 to happen, we term it a \emph{data
	stall}. Specifically, if training is blocked on Step 1, we call it a
fetch stall; the training is \emph{I/O bound} in this case.  Training
blocked due to Step 2 is termed prep stall; this causes the training
to be \emph{CPU bound}.  Data stalls cause the GPU to be idle, and
must be minimized to increase GPU utilization. 

The rate at which data items can be fetched from storage (Step 1)
depends primarily on the storage media. The rate at which data items
can be pre-processed (Step 2) depends upon the pre-processing
operations and the number of CPU cores available for pre-processing.

In general, if we prefetch data at rate $F$, 
pre-process it at rate $P$ and perform GPU computation
on it at rate $G$, then data stalls appear if $G > min(F, P)$, i.e.,
GPU processes data at a rate faster than it can be prefetched or 
pre-processed. 

Any fetch or prep stall implies idle GPU time, which must be minimized.
The fetch and prep stalls reported in this work
are unmasked
stall time;
i.e.,
the
stall time that shows up in the critical path,
inspite of being pipelined with compute. From now on, we call data prefetching simply \emph{fetch},
and pre-processing \emph{prep}.

	\section{Analyzing Data Stalls}
\label{sec-analysis}

To understand data stalls in DNN training and 
the fundamental reasons why data stalls exist, 
we perform a comprehensive analysis
on several DNNs by varying a number of factors,
such as the number of GPUs, GPU generation, the size of the DRAM
cache, the number of CPU threads etc. 

\subsection{Methodology}
\label{sec-methodology}
\vheading{Models and Datasets}. We analyze \textbf{nine}
state-of-the-art DNN models across three different tasks and four
different datasets as shown in Table~\ref{tbl-dataset-models}.  \revision{This
section focuses on the smaller ImageNet-1K dataset for image
classification models. Evaluation with large datasets like
ImageNet-22k and OpenImages is presented in
Section~\sref{sec-eval}}. The image and audio classification models are
taken from TorchVision~\cite{torchvision} and
TorchAudio~\cite{torchaudio} respectively; for object detection, we
use NVIDIA's official release of SSD300
v1.1~\cite{nvidia-ssd}.

\begin{table}[!t]
  \small
  \centering
  \ra{1.1}
    \mycaption{Models and datasets used in this work}{}
\vspace{-1.2em}
 \begin{tabular}{l@{\hskip5pt}r@{\hskip5pt}r@{\hskip5pt}}
	\toprule[1.2pt]
	     Task & Model & Dataset (Size)\\ 
	     \midrule
		\multirow{7}{*}{\shortstack[l]{Image \\Classification}}  & Shufflenetv2~\cite{zhang2018shufflenet} & \\
		& AlexNet~\cite{krizhevsky2012imagenet} & ImageNet-22k~\cite{imagenet-22k} \\
		& Resnet18~\cite{he2016deep} &  (1.3TB) \\
		& SqueezeNet~\cite{iandola2016squeezenet} & OpenImages-Extended \\
		& MobileNetv2~\cite{sandler2018mobilenetv2} &  ~\cite{kuznetsova2018open, sarincrowdsource} (645GB) \\ 
		& ResNet50~\cite{he2016deep} & Imagenet-1k~\cite{russakovsky2015imagenet} \\
		& VGG11~\cite{simonyan2014very} & (146GB)\\
	\midrule
	Obj Detection & SSD+Res18~\cite{liu2016ssd} & OpenImages~\cite{kuznetsova2018open} (561GB) \\
	\midrule
	Audio Classify & M5~\cite{dai2017very} & Free Music~\cite{defferrard2016fma} (950GB) \\
	\bottomrule[1.2pt]
   \end{tabular}

\label{tbl-dataset-models}
\end{table}

\vheading{\revision{Pre-processing}}.  \revision{For all DNNs, we use the same
pre-processing as in their original papers. More precisely, for the
image classification task, pre-processing includes image decoding, random crop, resizing to a fixed size,
and a random horizontal flip of the image. The object detection task performs a color
twist of the image, and a random crop and horizontal flip of the bounding box in addition to the
image transformations described for image classification. The audio model decodes and
down-samples input to 8kHz.}

\vheading{Training environment}. All experiments are performed on
PyTorch 1.1.0 using the state-of-the-art NVIDIA data loading pipeline,
DALI. We have empirically verified that DALI's performance is strictly
better than PyTorch, TF and MxNet's default data loaders; therefore we perform our analysis of data stalls using the strongest baseline, DALI.
We use two
distinct server configurations for our analysis as shown in
Table~\ref{tbl-analysis-sku}. Both these are part of a large production and research cluster
at Microsoft~\cite{philly,jeon2019analysis}, whose workload have guided the design
of several research systems for ML training~\cite{gandiva, tiresias, themis, gandiva-fair}.
These servers also closely resemble publicly available
cloud GPU SKUs~\cite{aws-p3, aws-p2}.
\confssd has configuration closest to
AWS p3.16xlarge~\cite{aws-p3} with gp2 storage~\cite{ebs}, while
\confhdd is closest to AWS p2.8xlarge~\cite{aws-p2} with st1
storage~\cite{ebs}.  Both our servers have 500GB DRAM, 24 physical CPU cores
, and 8 GPUs per server.

\begin{table}[!t]
  \small
  \centering
  \ra{1.1}
  \mycaption{Server configurations used}{We use two SKUs; each server has 24 CPU cores, 500\gb DRAM, and 8 GPUs.}
  \vspace{-1em}
 \begin{tabular}{cccccccc}
	\toprule[1pt]
	     \textbf{\emph{}} & GPU  & GPU  & Storage & Rand Read \\ 
	     \textbf{\emph{}} & Config & Mem(GB) & Media & (MBps)\\ 
	     \midrule
         \texttt{SSD-V100} & 8xV100 & 32 &  SSD & 530 \\
         \texttt{HDD-1080Ti} & 8x1080Ti & 11 & HDD & 15 -- 50 \\
	\bottomrule[1pt]
   \end{tabular}
\label{tbl-analysis-sku}

\end{table}

\vheading{Training parameters}. For experiments on \confssd, we use a
batch size of 512 per GPU for all image classification models, 128 per
GPU for SSD-Res18, 16 per GPU for M5 and perform weak scaling
for distributed training (while ensuring that the global batch size is
consistent with those widely used in the ML community). Since V100
GPUs have tensor cores, we use Apex mixed precision training with LARC
(Layer-wise Adaptive Rate Clipping), and state-of-the art learning
rate warmup schedules~\cite{goyal2017accurate}. On \confhdd, we use
the maximum batch size that fits the GPU memory (less than 256 for all
models) and perform full-precision training.

\vheading{Training metrics}. We run all the experiments presented here
for three epochs, and report the average epoch
time (or throughput in samples per second), ignoring the first epoch.
Since we start with a cold cache in our experiments, first epoch is
used for warmup. Measuring data stall
time does not require training to accuracy; per-epoch time
remains stable.

\subsection{Measuring data stalls using \tool}
\label{sec-tool}
We develop a standalone tool, \textit{\tool} that profiles data stalls
in DNN training.  Frameworks like PyTorch and TensorFlow provide an
approximate time spent on data loading and pre-processing per
minibatch, by simply placing timers in the training script. This is
insufficient and inaccurate for two reasons.  First, this technique
cannot accurately provide the split up of time spent in data fetch (from disk or
cache) and pre-processing operations. To understand if the training is
bottlnecked on \vio or CPU, it is important to know this split. Second,
frameworks like PyTorch and libraries like DALI use several concurrent
processes (or threads) to fetch and pre-process data; for
a multi-GPU data parallel training job, a data stall in one of the
data loading processes may reflect as GPU compute time for the other
processes, because all GPU processes wait to synchronize weight
updates at batch boundaries. Naively adding timers around data path
does not provide accurate timing information. Therefore, \tool uses a
differential approach.  \tool runs in three phases;
\begin{enumerate}[noitemsep,topsep=0pt,parsep=0pt,partopsep=0pt,leftmargin=18pt]
	
\item \vheading{Measure ingestion rate}.
  First, \tool pre-populates synthetic data at the GPUs and runs the job for a
  fixed number of epochs.
      This identifies the max data ingestion rate at the GPUs,
      with no fetch or prep stalls.
	\item \vheading{Measure prep stalls}. Next, \tool
	runs the training script with a subset of the given dataset,
	such that it is entirely cached in 
	memory, using all available CPU cores, and estimates the training speed. Since this run 
	eliminates fetch stalls, any drop in throughput compared
	to (1) is due to prep stalls.
	\item \vheading{Measure fetch stalls}. Finally,
	\tool runs the training script by clearing all caches, 
	and setting maximum cache size to a user-given limit, to
	account for fetch stalls. The difference between (2)
	and (3) is the impact of fetch stalls.
\end{enumerate}


\subsection{Data Stalls in DNN Training}
\label{sec-data-stall}
Our analysis aims to answer the following
questions:
\vspace{-1.5em}
\setlength\mylength{\dimexpr.5\columnwidth-10\tabcolsep-0.5\arrayrulewidth\relax}
\newcolumntype{M}[1]{>{\centering\arraybackslash}m{#1}}
\begin{table}[!h]
    \small
 	\centering
 	\ra{0.95}
 	 \begin{tabular}{!{\VRule[1pt]}M{0.09\textwidth}!{\VRule[1pt]}m{0.3\textwidth}!{\VRule[1pt]}m{0.04\textwidth}!{\VRule[1pt]}}
 	\specialrule{1.2pt}{0pt}{0pt}
 	 \rowcolor{white}
 	
 		\vheading{ Fetch Stalls (Remote)} & Is remote storage
 	   a bottleneck for training? &   \sref{sec-fetch-remote}\\
         \specialrule{0.5pt}{0pt}{0pt}
 		\vheading{Fetch Stalls (Local)} & When does the local storage device
 		(SSD/HDD)
 		become a bottleneck for DNN training?  &  \sref{sec-fetch}\\
 		 \specialrule{0.5pt}{0pt}{0pt}
 	   \vheading{Prep Stalls} &  When does data prep at the CPU 
 	   become a bottleneck for DNN training? & \sref{sec-prep} \\
 		 \specialrule{0.5pt}{0pt}{0pt}
 	   \vheading{Generality} & Do fetch and prep stalls exist in other training platforms like TensorFlow? &  \sref{sec-tf}\\
 		 \specialrule{1.2pt}{0pt}{0pt}
 	\end{tabular}
\label{tbl-analysis-q}
\vspace{-1.5em}
\end{table}

\subsubsection{\textbf{When dataset resides on remote storage}}
\label{sec-fetch-remote}
\begin{figure}[!t]
  \centering 
   \includegraphics[width=.48\textwidth]{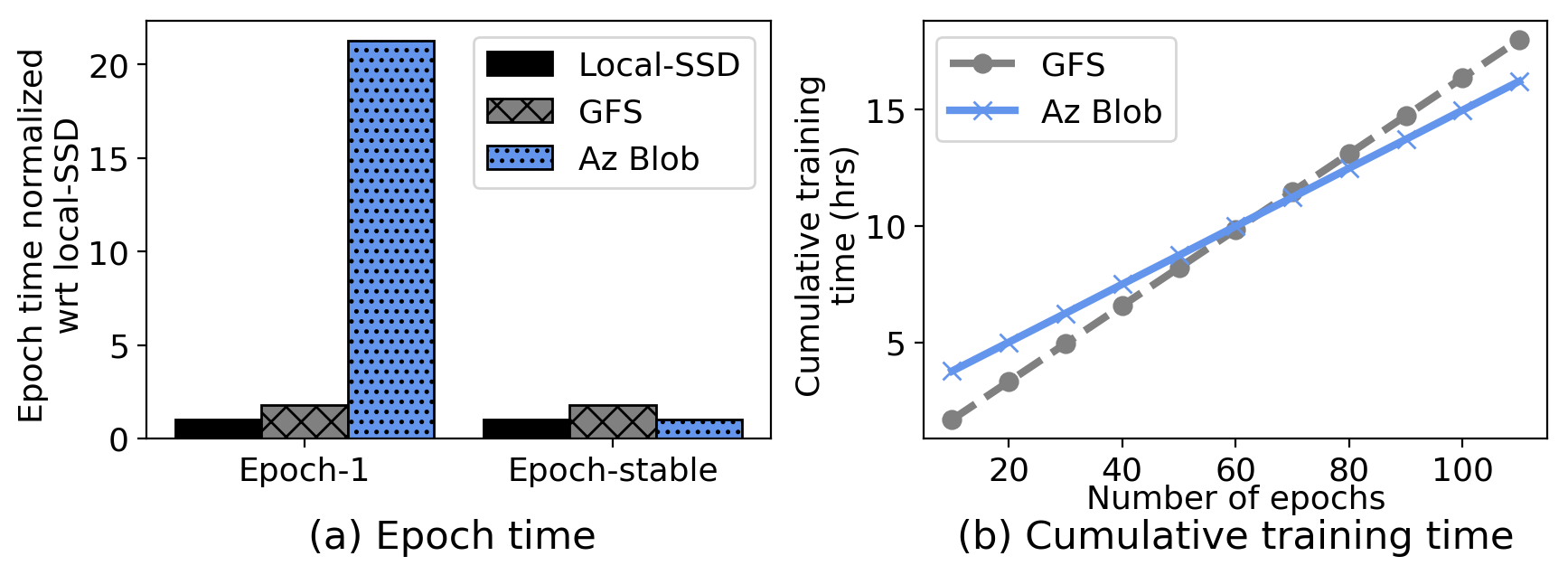}
    \vspace{-2.4em}
  \mycaption{Training with remote stores}{The high download cost of blob is amortized over training for a large \# of epochs}
  \vspace{-1.6em}
  \label{fig-remote}

\end{figure}

Datasets used for training DNNs could reside locally on the persistent storage of a 
server, or on shared remote storage such as distributed file systems (HDFS, GlusterFS - GFS),
or object stores (S3, Azure blobs). We analyze the impact of two kinds of remote backends;
a distributed file system, GlusterFS (GFS) and the Azure blob object store accessed via blobfuse. 
When data resides remotely, the first epoch of training fetches data
over the network and stores it locally for subsequent use. Cluster
file systems like GFS use the OS Page Cache to speed up subsequent
accesses. Blobfuse downloads the dataset on to local SSD, and mimics
local training from the second epoch.  Figure ~\ref{fig-remote}a
compares the epoch time for ResNet18 on \confssd using GFS, blobfuse,
and local SSD for the first epoch and a stable-state epoch with warmed
up cache.

The data stall overhead of BlobFuse is especially high in the first
epoch when it downloads the entire dataset to local storage, and can
result in 20\myx higher training time as compared to GFS. 
Unsurprisingly, during the steady state epochs, data stall
overheads when using the local SSD and BlobFuse are similar (as the
blob data is cached on the local SSD); GFS results in more data stalls
as it validates metadata of cached data items over the network every
time a data item is accessed. Blobfuse does not incur any network
cost beyond first epoch, if the dataset fits on local SSD.

As shown in Figure~\ref{fig-remote}b, for the ImageNet1K dataset,
for BlobFuse, the cost of downloading the entire dataset in the first
epoch is amortized as we train for a longer number of epochs, making
the remote Blobstore a better fit compared to GFS when models are trained to accuracy
for over 60 epochs.

Athough datasets are growing in size, large datasets that are
publicly available fit entirely on local storage (but not in memory)~\cite{training-big,
	abu2016youtube, kuznetsova2018open, defferrard2016fma, imagenet-22k, russakovsky2015imagenet}.
Therefore, a common training scenario is to pay a one-time download
cost for the dataset, and reap benefits of local-SSD accesses
thereafter (default and recommended mode in the Microsoft
clusters). Therefore, in the rest of the work, we analyze fetch stalls
in scenarios where dataset is present locally on a server, but is not
entirely cached in memory.

\subsubsection{\textbf{When datasets cannot be fully cached}}
\label{sec-fetch}
ML-optimized cloud servers with 500GB DRAM can only cache 35\%
of ImageNet-22K, or 45\% of the FMA dataset, or 65\% of the OpenImages
dataset, although they entirely fit on local storage. 
Popular datasets like ImageNet-1K cannot be fully cached on
commonly used cloud SKUs like AWS \vtt{p3.2xlarge}, which has 61 \gb
DRAM. When datasets don't fit in memory, and the fetch rate($F$)
$<$
compute rate ($min(P,G)$), fetch stalls occur.

\begin{figure}[!t]
  \centering 
   \includegraphics[width=.45\textwidth]{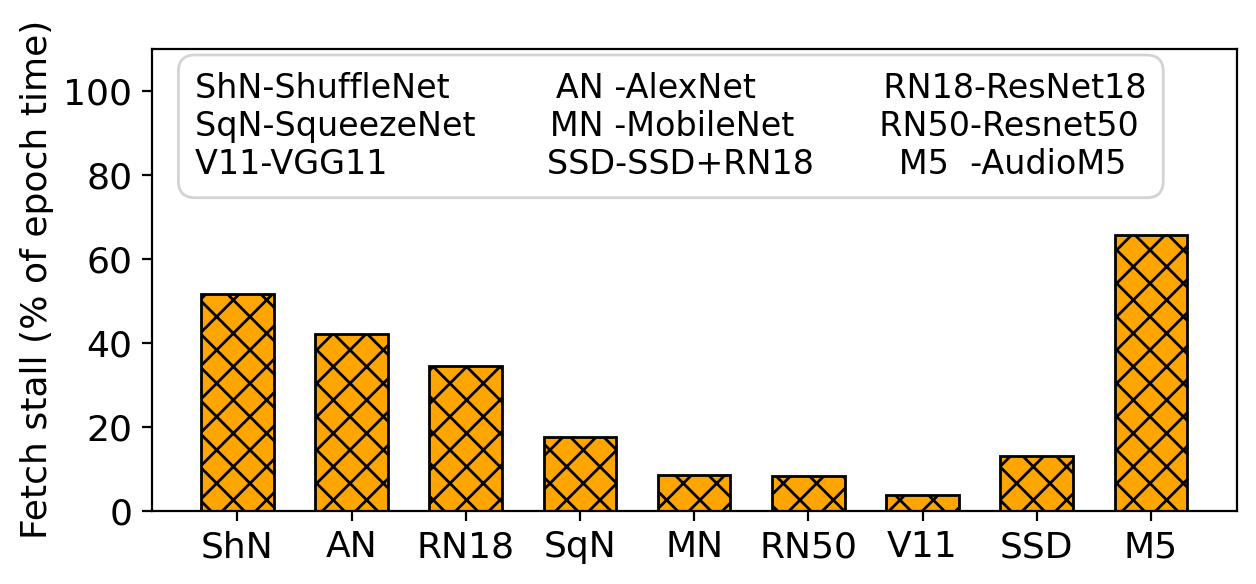}
  \vspace{-1em}
   \mycaption{Fetch stalls}{
     Several DNNs experience significant stalls waiting for \vio, when
     training on \confssd with 35\% of their dataset cached.
   }
  \vspace{-1em}
  \label{fig-analysis-cache-dnn}

\end{figure}

\vheading{Fetch stalls are common if the dataset is not fully cached in memory}.
Figure~\ref{fig-analysis-cache-dnn} shows the percentage of per 
epoch time spent on \vio for nine different DNNs when 35\% ( for \eg ImageNet-22k on 500GB server) 
of their respective
datasets can be cached in memory on \confssd. DNNs spend 10 --70\% of their epoch
time on blocking \vio, despite pipelining and prefetching, simply
because the compute rate is higher than fetch rate.

\begin{figure}[!t]
  \centering 
    \includegraphics[width=.48\textwidth]{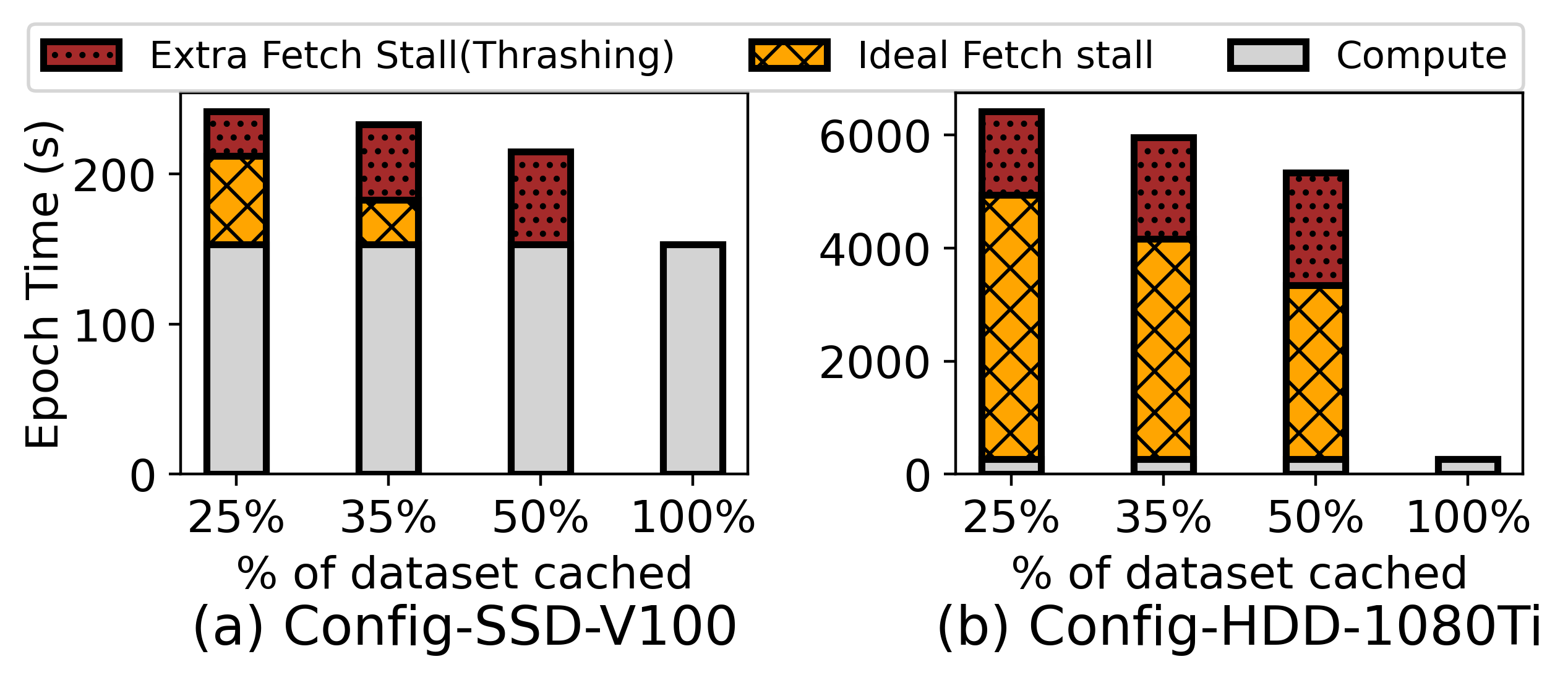}
    \vspace{-2.5em}
  \mycaption{ResNet18 with varying cache}{This stacked bar chart splits epoch time into time spent in compute, ideal fetch stalls, and the additional fetch stall due to thrashing.}
  \vspace{-1.2em}
  \label{fig-analysis-cache-res18}

\end{figure}

\vheading{OS Page Cache is inefficient for DNN
  training}. DNN training platforms like PyTorch, TensorFlow and
libraries like DALI, rely on the operating system's Page Cache to
cache raw training data in memory. Unfortunately, the OS Page Cache
leads to thrashing as it is not efficient for DNN training. \emph{If 35\% of
the data can be cached, then an effective cache should provide 35\%
hits; instead, the Page Cache provides a lower hit rate}. For a 146
\gb data set, each epoch should see only 65\% of the dataset, or 95\gb,
fetched from storage. Instead, we observe 85\% of the dataset fetched
from storage every epoch; the 20\% difference is due to
thrashing. Figure~\ref{fig-analysis-cache-res18} shows the fetch
stalls, including those due to thrashing, when using PyTorch with
DALI. An effective cache for DNN training must eliminate thrashing to
reduce fetch stalls to the minimum shown in
Figure~\ref{fig-analysis-cache-res18}.

\vheading{Lack of coordination among caches leads to
  redundant \vio in distributed training}. In distributed training
jobs, the data to be fetched and processed is divided randomly among
servers, and changes every epoch. As a result,
each server often has to fetch data from storage every epoch; this is
done even if the required data item is cached in an another server that is a part
of the distributed training job. This lack of coordination among
caches makes distributed training storage \vio-bound. 
When training Resnet50 on ImageNet-1K (146\gb) across
two servers having a total cache size of 150\gb, each server fetches
45\gb from storage in each epoch (despite the fact that the 
other server might have this data item in its cache). 
On \confhdd, this leaves ResNet50 stalled on \vio for 75\% of its epoch time.


\vheading{Lack of coordination in HP search results in
  redundant \vio}. HP search is performed by launching several
parallel jobs with different HP on all available GPUs in
a server~\cite{liaw2018tune}. All HP jobs access the same dataset in a
random order in each epoch, resulting in cache thrashing and read
amplification.
When 8 single-GPU jobs are run in a server (35\% cache), there is 7\myx read amplification per
epoch (884 \gb read off storage compared to 125 \gb for one job), 
which slows down HP search on ResNet18 by 2\myx on \confssd.

\begin{figure*}[!htb]
\minipage[b]{0.39\textwidth}
  \includegraphics[width=\linewidth]{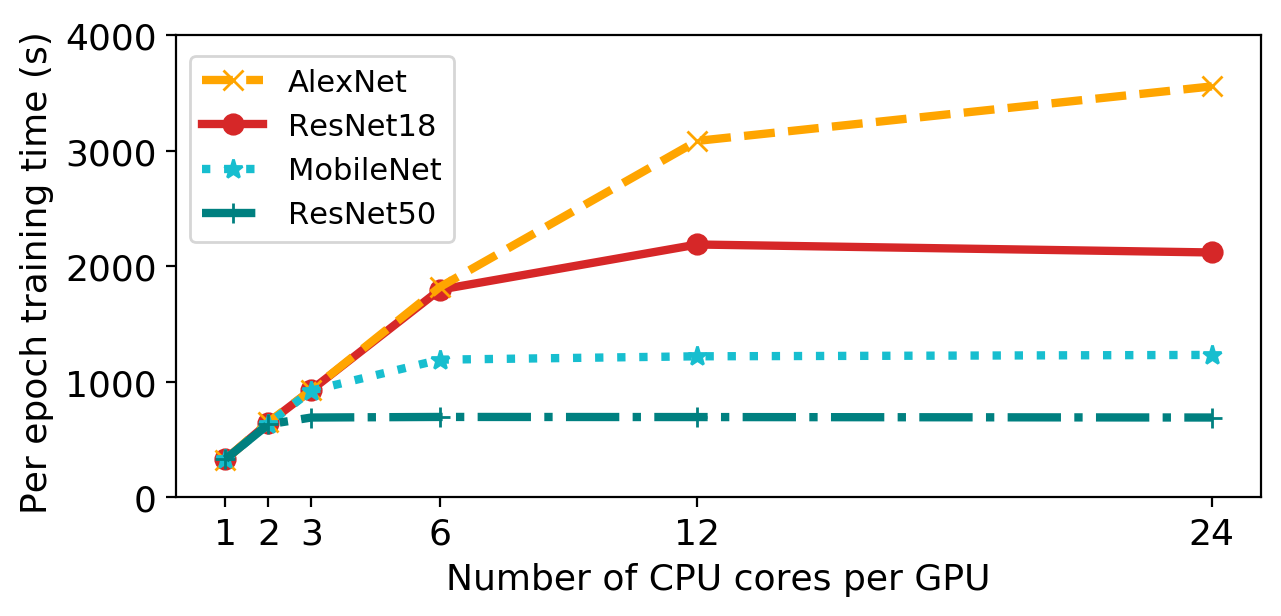}
     \vspace*{-0.75cm}
  \caption{Impact of CPU cores on training}\label{fig-analysis-prep-cpu}
\endminipage\hfill
\minipage[b]{0.37\textwidth}
  \includegraphics[width=\linewidth]{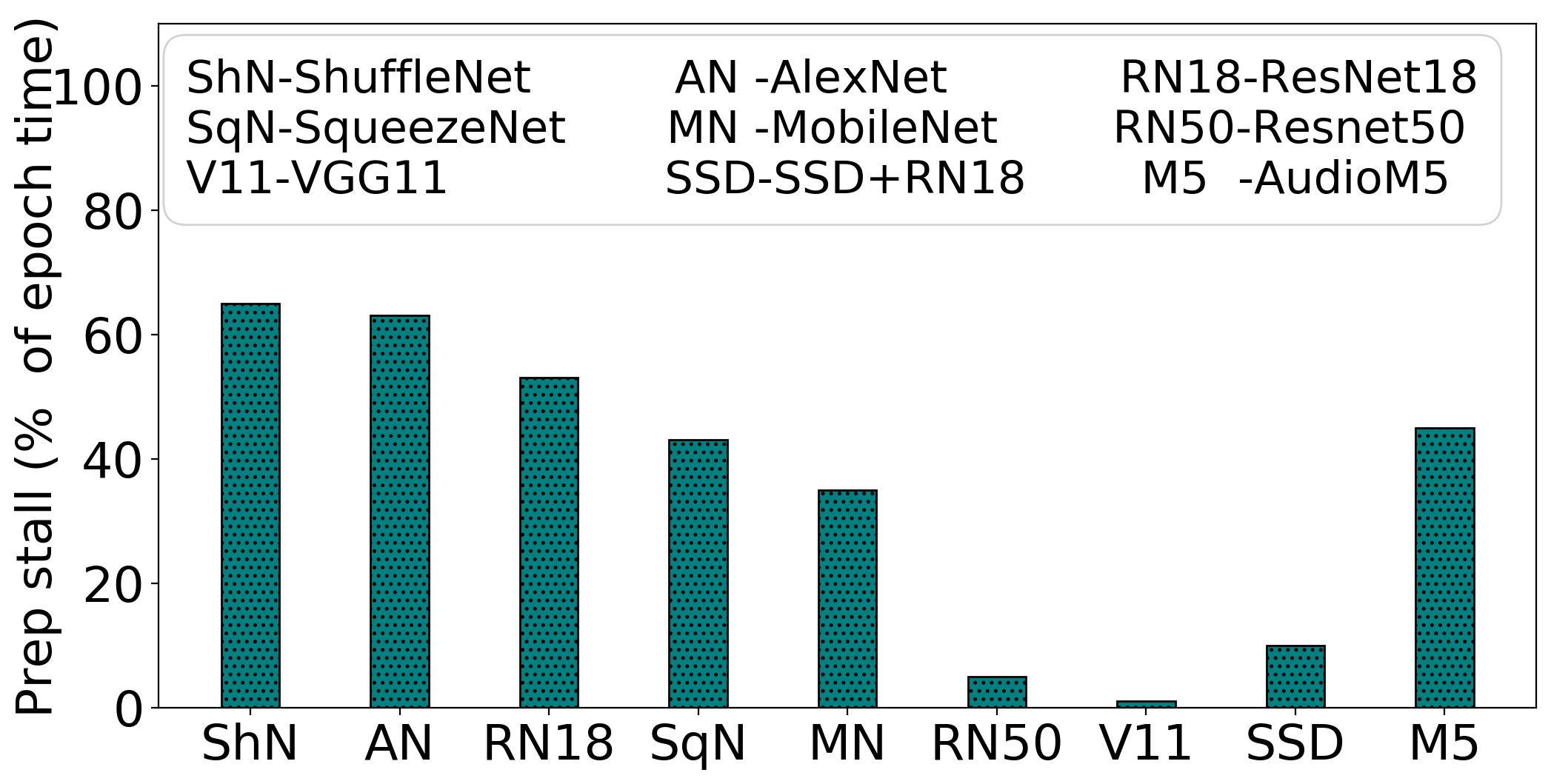}
     \vspace*{-0.75cm}
  \caption{Prep stall across DNNs}\label{fig-analysis-prep-all}
\endminipage\hfill
\minipage[b]{0.23\textwidth}%
  \includegraphics[width=\linewidth, height=3.4cm]{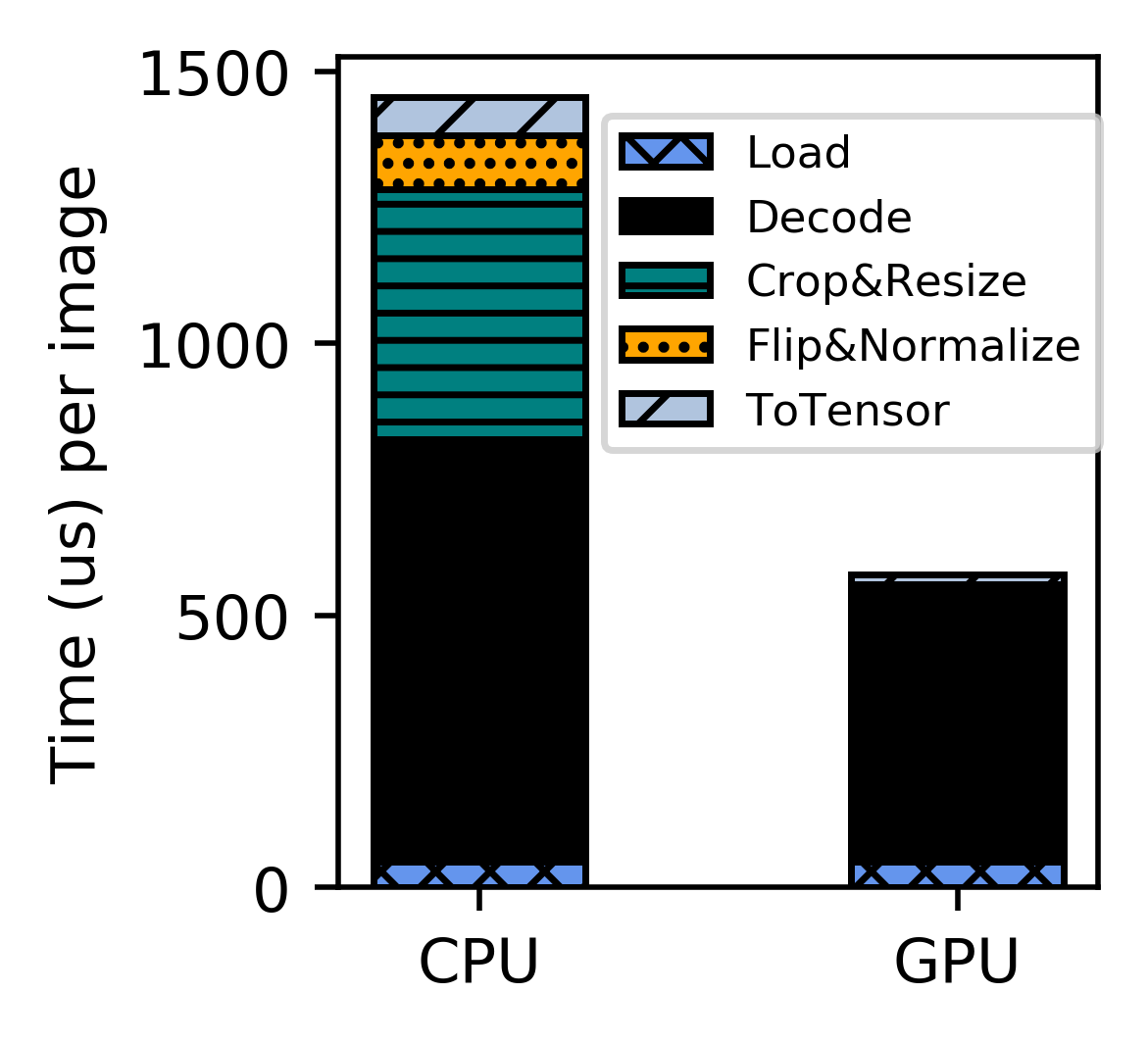}
   \vspace*{-0.75cm}
  \caption{Breakdown prep}\label{fig-prep-breakdown}
\endminipage
\vspace*{-1em}
\end{figure*}

\begin{figure}[!t]
  \centering 
   \includegraphics[width=.48\textwidth]{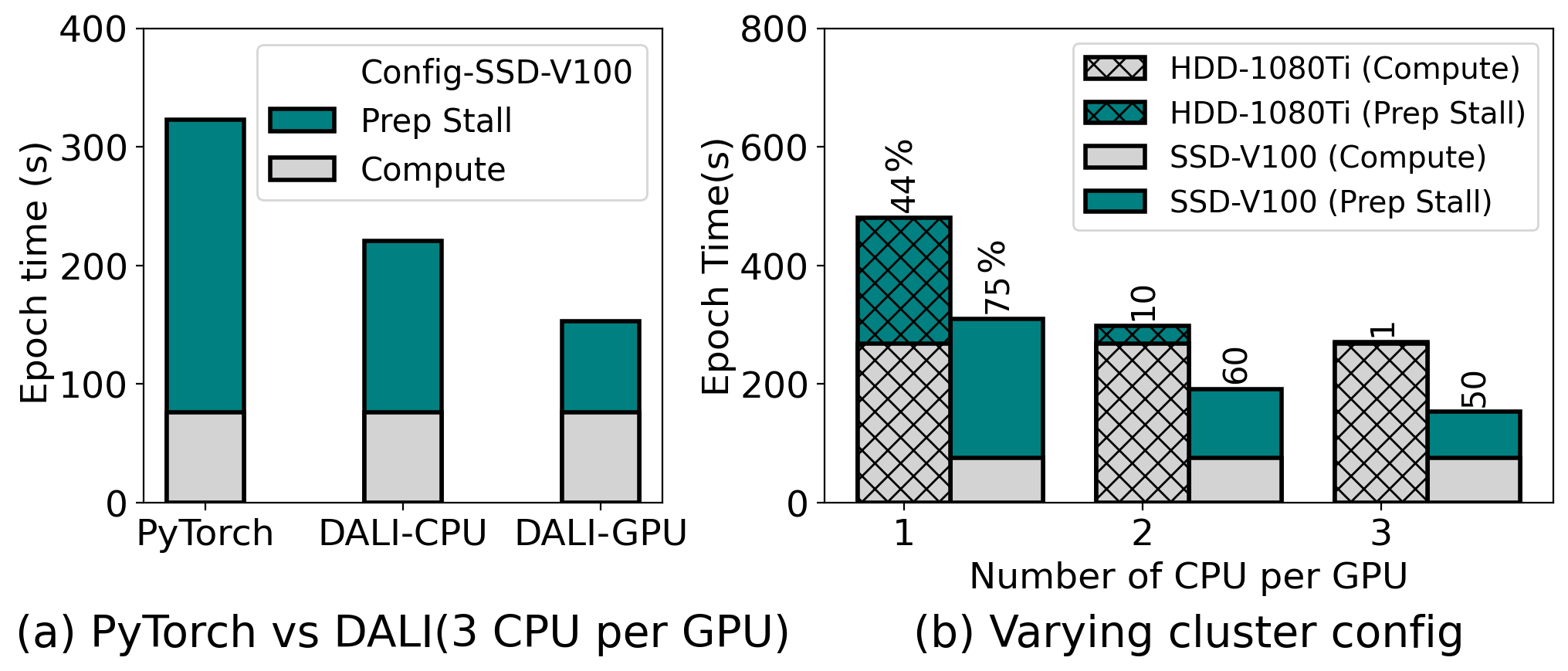}
    \vspace{-2em}
  \mycaption{8-GPU ResNet18 training}{Even with DALI, faster GPUs like V100 have upto 50\% prep stalls.}
  \vspace{-1em}
  \label{fig-analysis-prep-res18}

\end{figure}

\subsubsection{\textbf{When datasets fit in memory}}
\label{sec-prep}

We now analyze the impact of CPU pre-processing on DNN training in the
scenario where the entire dataset is cached in memory of a single server, thus
eliminating fetch stalls due to storage \vio.

\vheading{DNNs need 3--24 CPU cores per GPU for
  pre-processing}.  Figure~\ref{fig-analysis-prep-cpu} shows how DNN
training throughput changes as we vary the number of CPU
pre-processing threads (per V100 GPU) for four models. For computationally
complex models like ResNet50, 3 -- 4 CPU cores per
GPU is enough to prevent prep stalls; for computationally lighter models
like ResNet18 or AlexNet, as many as 12 -- 24 CPUs per GPU are needed
to mask prep stalls. 
Even on NVIDIA's AI-optimized DGX-2,
there are only three CPU cores per GPU; thus, many models have prep
stalls on the DGX-2 (~\sref{sec-dgx2})

\vheading{DALI is able to reduce, but not eliminate prep
  stalls}. DALI \revision{has a GPU-based prep mode to offload a part of pre-processing to the GPU}.
As shown in
Figure~\ref{fig-analysis-prep-res18} (a), when all pre-processing except decoding is offloaded to the
GPU for training ResNet18, prep stalls reduce. The effectiveness of DALI
depends on the GPU speed; for example, on the slower 1080Ti, DALI is
able to eliminate prep stalls using three CPU threads per GPU. On
the faster V100 though, DALI still results in 50\% prep stalls when
using three CPU cores per
GPU, and the GPU for pre-processing.  \revision{Since DALI cannot automatically split pre-processing between CPU and GPU, we empirically find that
offloading all operations except decoding to the GPUs offered best performance for the object detection and image classification models except ResNet50 and VGG11. These two models are computationally complex and slow down 
if pre-processing is performed at the GPU. Furthermore, DALI does not support audio decoding at the GPU. Therefore, pre-processing is entirely done at the CPU for ResNet50, VGG11 and audio M5 models.} Figure~\ref{fig-analysis-prep-all} shows that prep stalls exist across different DNNs when training with eight GPUs each with 3 CPUs \revision{when using the best of CPU or GPU-based prep mode of DALI as described above.}

\vheading{Decoding is very expensive!}
To understand what makes prep expensive, we
run a microbenchmark to breakdown the cost of different stages
in a typical prep pipeline for image classification tasks.
An input image is first loaded into memory, decoded, and then randomly
transformed (crop, flip), and finally copied over to the GPU as a tensor that
can be processed. Fig ~\ref{fig-prep-breakdown} shows the time 
taken for each operation when prep is done on CPU and GPU by DALI. 
We see that offloading 
prep to the GPU provides significant speedup at the expense of GPU memory usage (+5GB!). 
Second, a majority of time during prep is spent in decoding images.

\begin{figure}[!t]
  \centering 
   \includegraphics[width=.4\textwidth]{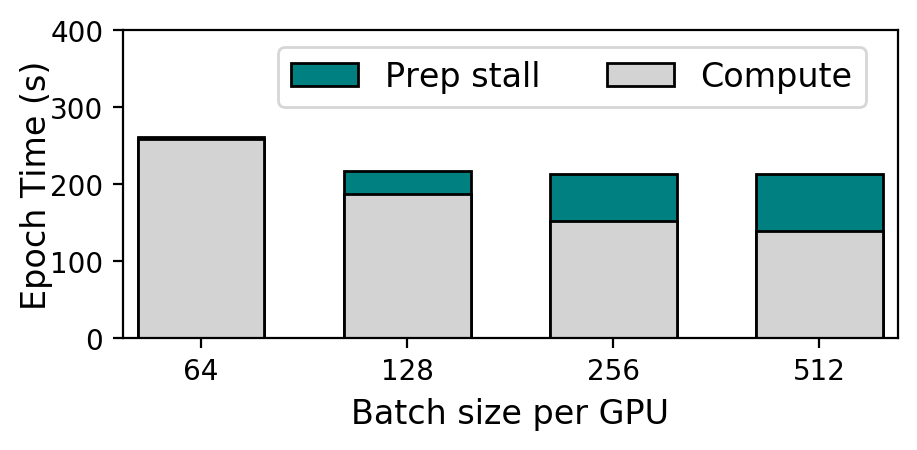}
  \vspace{-1.5em}
  \mycaption{\revision{Impact of batch size on prep}}{}
  \label{fig-analysis-batch}
  \vspace{-2em}
\end{figure}

\vheading{Impact of batch size}. 
\revision{The impact of batch size on GPU computational efficiency is well
studied [47, 89]; larger batch sizes utilize the massive GPU parallelism better, 
and also reduce the number of weight updates
(inter-GPU communication) per epoch, resulting in faster training. 
Figure~\ref{fig-analysis-batch} shows the impact of varying the batch size on epoch
time and the percentage of epoch time spent on prep stalls for
MobileNetv2. As computational efficiency increases with larger
batches, training becomes CPU bound due to data prep. Note that,
although the required GPU compute time dropped with a larger
batch size, per epoch time remained same due to prep stalls. This
graph makes an important point; as compute gets faster (either due
to large batch sizes, or the GPU getting faster), data stalls squander the
benefits due to fast compute. }

\setlength\mylength{\dimexpr.5\columnwidth-10\tabcolsep-0.8\arrayrulewidth\relax}

\begin{figure}[!t]
	\centering
	\ra{0.95}
	\begin{tabular}{l}
		\kern-0.8em\subfloat[100\% Cache \label{fig-analysis-framework-prep}]{\adjustbox{raise=-2.6pc}{\includegraphics[width=2.5cm]{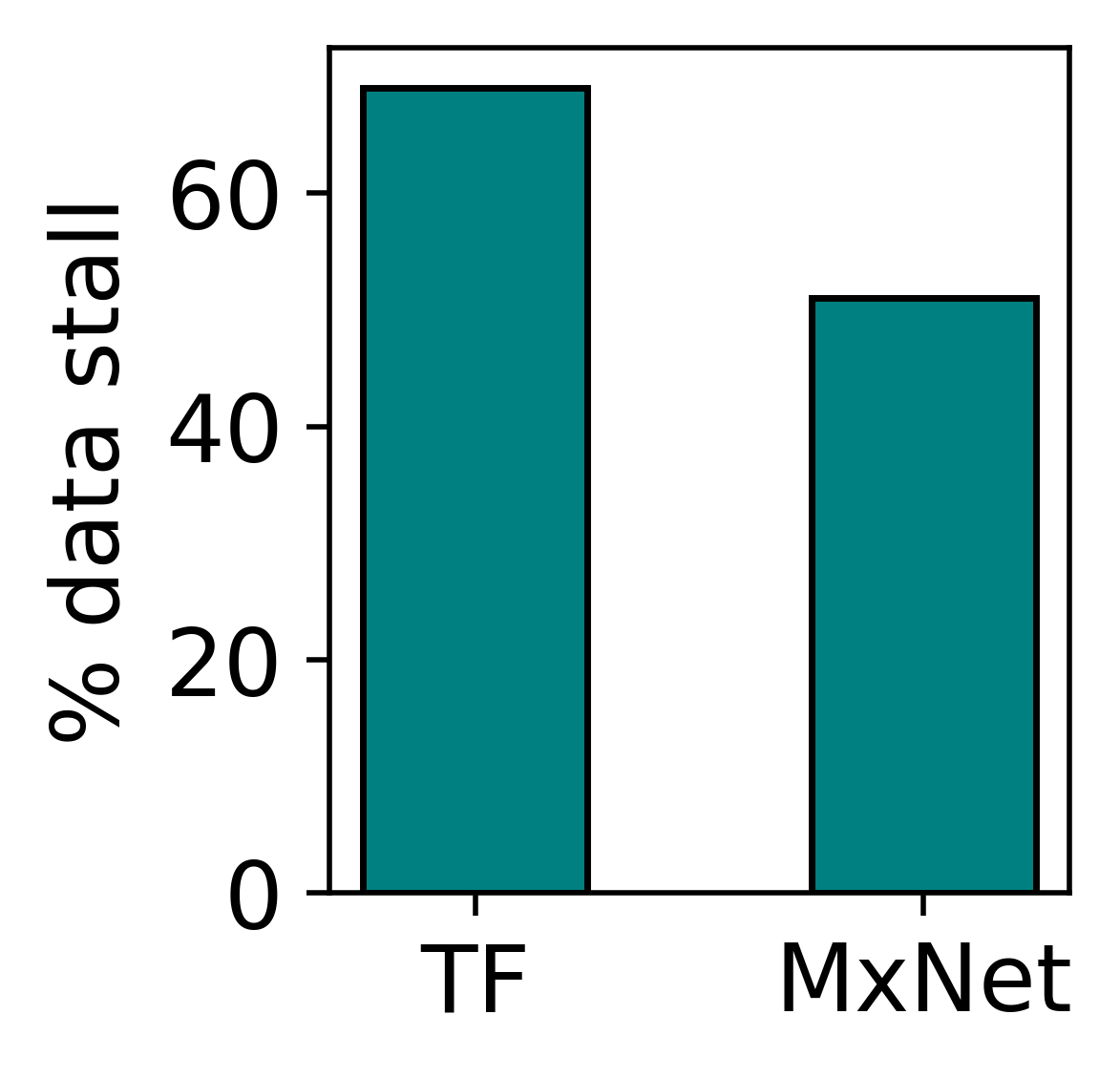}} }
		\quad
		\kern-0.8em\subfloat[Varying cache (TensorFlow)
		\label{tbl-analysis-tf-small}]{
		 \begin{tabular}{@{}c|c|cc@{}}
			\toprule[1.2pt]
			\% cached  &   8-GPU job & \multicolumn{2}{c}{8-job HP}\\
			(Sz:146GB) &  Cache Miss	   &   \vio(GB) & Read amp \\
			\midrule
			50\% & 91\% & 860 &  6.14\myx \\
			35\% & 94\% & 1010 & 7.21\myx \\
			25\% & 97\% &  1019 & 7.28\myx \\
			\bottomrule[1.2pt]
		\end{tabular}
		}%
		\label{fig-analysis-frameworks}
	\end{tabular}
\vspace{-1em}
	\caption{Data stalls across frameworks}
	\vspace{-1.5em}
\end{figure}

\vheading{Redundant pre-processing in HP search results
  in high prep stalls}.  During HP search, concurrent jobs process the
same data. Currently, there is no coordination; if there are 8 HP
jobs, the same data item is processed eight times. This is made worse
by the fact that all HP jobs share the same set of CPU threads,
leading to fewer CPU threads per GPU, and higher prep
stalls. When 8 single-GPU ResNet18 HP jobs run on \confssd, each job gets 3 CPU
for prep and incurs a 50\% prep stall as shown in  
Figure~\ref{fig-analysis-prep-all}. Coordinating these HP search jobs on a single 
server can potentially eliminate prep stalls, as all available CPU
(24 cores) can be used to prep the dataset exactly once per epoch and reused across jobs
(Figure~\ref{fig-analysis-prep-cpu} shows ResNet18 requires 12 CPUs per GPU to
eliminate prep stalls).

\subsubsection{\textbf{Data stalls exist across training frameworks}}
\label{sec-tf}

To generalize our findings on data stalls across different training
platforms and data formats, we analyze the prep and fetch stalls in
TensorFlow (TF) using the binary TFRecord format, and MxNet using RecordIO format. 
Unlike
PyTorch, TF does not store training data as small individual
raw files. Instead, it shuffles the small random files, serializes
it, and stores them as a set of files (100-200MB each) called
TFRecords. TFRecords make reads more sequential. MXNet also use a similar serializing technique for data called
RecordIO~\cite{recordio}.

Figure~\ref{fig-analysis-framework-prep} shows that both native
TF and MxNet spend 65\% and 50\% of the epoch time on prep stall
for a 8-GPU ResNet18 training job when the dataset is entirely
cached in memory.
Next, Table~\ref{tbl-analysis-tf-small} shows the percentage of misses in
the Page Cache for a 8-GPU training job and the \vio amplification due
to lack of coordination in HP search for varying cache sizes in TF. 
TFRecord format results in 40\%
higher cache misses than the ideal because, the sequential access
nature of TFRecords (and RecordIO) is at odds with LRU cache replacement policy of
the Page Cache, resulting in a pathological case for LRU
(this is 20\%higher than PyTorch).  
The lack of co-ordination
in HP jobs results in upto 7.2\myx read amplification; although
all jobs read the same 140 \gb dataset, the total disk \vio was 1.1
TB.

\subsubsection{\textbf{Analysis of NLP Models}} \revision{In addition to
vision and audio models in table~\ref{tbl-dataset-models}, we evaluated data stalls on two
language models; Bert-Large pretraining~\cite{devlin2018bert} on Wikipedia \&
BookCorpus dataset~\cite{zhu2015aligning} for language modeling and
GNMT~\cite{wu2016google} on WMT16~\cite{wmt16} (EN-De) dataset for
translation, which do not exhibit data stalls.
Language models have a distinct pre-processing regime compared to  vision-based models. 
Text-based models pre-process and shuffle the data corpus \emph{once} before training, and then 
reuse it every epoch. There is no online, random, heavy-weight  pre-processing performed in every epoch
unlike vision models. 
Therefore, the language models do not exhibit
data stalls in our training environment. 
However, data stalls may show up in these models if
GPUs get faster or the computation requirements for these models gets
lower due to compact model representations.}

	\section{\tool : Predictive analysis}
While all the experiments in \sref{sec-data-stall} are run on physical
servers, we extend \tool to help a user simulate these experiments
without having to run all different configurations on physical
servers.  
While there exists prior work that predict the performance of
a DNN, they focus on profiling the layer-wise
performance of DNN~\cite{tpu-tools, mxnet-profile},
low level perf counters for accelerators~\cite{nvidia-profile, iyer2016gpu},
or finding optimization opportunities at the neural network 
layer level~\cite{zhu2020daydream}. 
In contrast, \tool analyzes the
implication of CPU, memory, and storage
on the performance of a DNN and answers what-if questions.

This is a powerful means of analyzing whether
throwing more hardware at the problem will solve the issue of data
stalls. For instance, if training is dominated by fetch stalls
(bottlenecked on disk bandwidth), then increasing the number of CPU
cores on the machine has no benefit; either DRAM capacity has to be
increased, or the disk must be replaced with a higher bandwidth
one. Similarly, if the training job is bottlenecked on prep,
then increasing DRAM has no effect on training time.  \tool is useful
in scenarios like this, to predict the performance of a model as we
scale up CPU, memory, or storage.

\subsection{Estimating data stalls}
Consider the different components involved in a typical DNN  
data pipeline as in Figure~\ref{fig-rate}; 
data is fetched from cache (and store) with an effective 
prefetch rate $F$, pre-processed at the CPU at a rate $P$ and
processed at the GPU at a rate $G$. 
To perform predictive analysis, \tool measures 
several metrics related to the data pipeline of the model;
the maximum ingestion rate at the GPU $(G)$, the rate 
of CPU prep $(P)$, the
rate of cache fetch $(C)$, and the rate of storage fetch $(S)$. 
Using these metrics,
\tool models the training iteration 
to answer what-if questions such as, how much
DRAM cache is required for this model to
eliminate fetch stalls?; 
\tool 
collects these metrics for a model as follows.

\vheading{(i) Measure ingestion rate $(G)$}. To find the
maximum possible speed at which the DNN can train, 
\tool first runs the job script for a fixed number iterations (default:100)
with synthetic data that is pre-populated at the GPUs.
It then calculates $G$ as,
\vspace{-0.5em}
\begin{equation}
G = \frac{\mbox{Total samples processed in (i)}}{\mbox{Time for (i)}} \\
\vspace{-1.5em}
\label{eq-gpu}
\end{equation}

\begin{equation}
\mbox{Samples processed} = \mbox{\#iterations} \times \mbox{global batch size}\\
\end{equation}

\vheading{(ii) Measure prep rate $(P)$}. Next, \tool
executes the training script with the given dataset
by ensuring that the subset of data used is cached in 
memory, using all available CPU cores. Additionally,
the GPU computation is disabled to only run the 
data loader. This is required because, if $P \geq G$,
then we cannot measure P using the knowledge of
runs (i) and (ii), as prep will be pipelined
with GPU compute. Therefore, \tool disables
GPU computation and estimates P in the same way
as Eq~\eqref{eq-gpu}.

\vheading{(iii) Measure storage fetch rate $(S)$}.
Storage fetch rate is the maximum random
read throughput of the storage device. To measure
this, \tool runs the data loader (with a cold cache,
disabling both pre-processing and GPU compute), with
all CPU cores.

\vheading{(iv) Measure cache fetch rate $(C)$}.
To measure the rate at which data can be fetched 
from cache, \tool uses a microbenchmark to 
measure memory bandwidth and uses it as an 
approximation for $C$. Note that run (ii) actually
includes the time to fetch cached items as well;
however we see that the cache fetch rate is very high 
(few tens of GBps), and does not add noise to the
measurement of prep rate.


\begin{figure}[!t]

  {\includegraphics[width=0.48\textwidth]{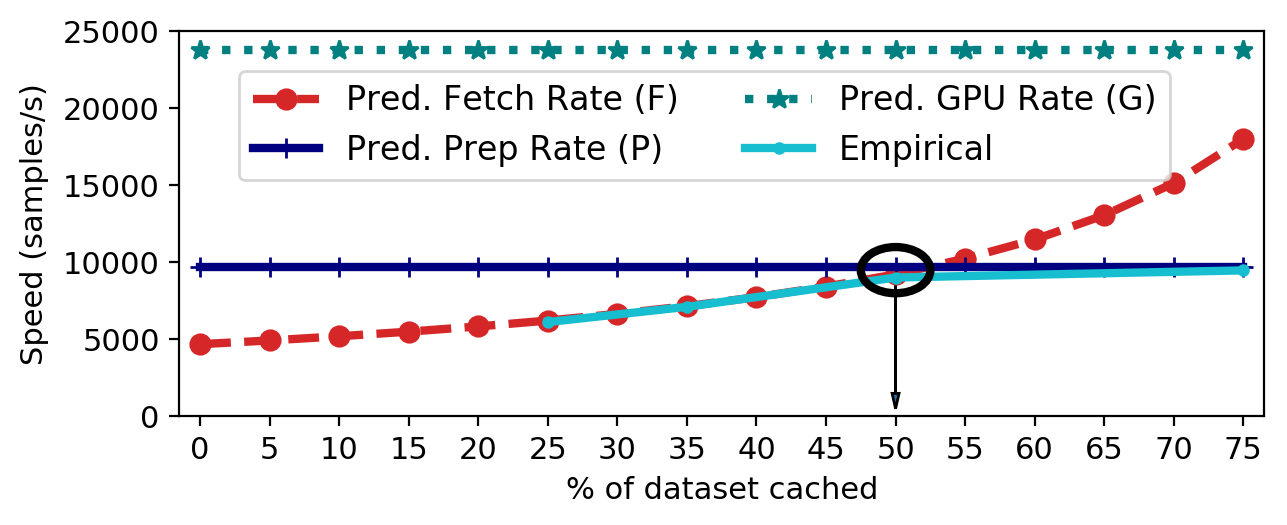}}
   \vspace{-2.5em}
  \mycaption{Estimating optimal cache size with \tool}{ }
  \label{fig-tool-cache}
  \vspace{-1em}
\end{figure}

\subsection{Example : Predicting optimal cache size}
We now describe an example of what-if analysis with
\tool. We show how \tool answers the question : 
\emph{how much DRAM cache does the DNN need to eliminate 
	fetch stalls?}

To predict the implication of cache size, \tool 
calculates the effective prefectch rate $(F)$ for 
a given cache size $(x$ \% of the dataset). Here, we assume that
the cache implements an efficient policy like
\minio; i.e., a cache of size $x$ items has atleast
$x$ hits per epoch.

$F$ is computed as follows. Say the size of the dataset is
$D$ samples, and cache is $x\%$ of the dataset. Therefore,
in an epoch, the total time to read the dataset is given by
\vspace{-0.4em}
\begin{equation}
T_f = \frac{D \times x}{C} + \frac{D \times (1-x)}{S} \\
\label{eq-fetch-time}
\end{equation}

The fetch rate is then calculated as,
\begin{equation}
F = \frac{D}{T_f} = \frac{D}{\frac{D \times x}{C} + \frac{D \times (1-x)}{S}}\\
\label{eq-fetch}
\end{equation}

Since $C >> S$, $F \propto \frac{1}{1-x}$, i.e, the effective fetch
rate increases, as the number of uncached items per epoch decreases.
Since \tool has already estimated values of $D$, $C$, and $S$, given a 
cache percentage $x$, \tool can predict the fetch rate using Eq~\eqref{eq-fetch}.

To evaluate how accurately \tool can 
answer this question, we run the actual experiment by
varying cache size on a physical server (empirical), and 
comparing it to the predictions of \tool 
for AlexNet on \confssd with Imagenet-1K.  
 Figure~\ref{fig-tool-cache} plots the predicted 
values of $F$, $P$, and $G$, alongside empirical speed while
varying cache size. First, we observe that  the predicted training speed
($min(F,P,G)$) is a maximum of
4\% off the empirical results. Second, using these predictions, \tool
can estimate the optimal cache size for the model 
by comparing it with prep rate (P) and GPU ingestion rate (G). 
To eliminate fetch stalls, $F > min(P,G)$ as shown by the intersection 
in Figure~\ref{fig-tool-cache}.  At lower cache sizes,
training is \vio bound, however, a cache that is 
50\% of the dataset size is sufficient to eliminate fetch stalls;
larger cache (more DRAM) is not beneficial beyond this point,
as training becomes CPU-bound. 
A comprehensive list of data pipeline rates ($G, P, F$) for several models, datasets, 
and configurations is in the Appendix (\sref{sec-app-rates}).


	\section{Mitigating Data Stalls}
\label{sec-design}

Based on the insights from our analysis,
we explore ways of mitigating data stalls using domain specific
techniques that reduces cache misses,
and eliminates redundancy in data fetch and prep.
We further discuss how to reduce the cost of decoding
in future work(~\sref{sec-disc}).

\vspace{-1em}
\setlength\mylength{\dimexpr.5\columnwidth-12\tabcolsep-1\arrayrulewidth\relax}
\newcolumntype{M}[1]{>{\centering\arraybackslash}m{#1}}
\begin{table}[!h]
  \small
  \centering
  \ra{0.8}
 \begin{tabular}{M{0.1\textwidth}M{0.25\textwidth}M{0.09\textwidth}}
	\toprule[1.2pt]
    \textbf{\emph{Technique}} & \textbf{\emph{Impact}} & \textbf{\emph{Benefits}}\\
    \midrule
    \vheading{\minio Cache}& DNN-aware caching to minimize  IO by reducing cache misses per epoch (~\sref{sec-minio}) & Single-server training\\
    \midrule
    \vheading{Partitioned \minio Cache} & Eliminate redundant fetch by coordinating remote \minio caches (~\sref{sec-part-cache}) & Distributed training\\
    \midrule
    \vheading{Coordinated Prep}& Eliminates redundant fetch and prep across jobs (~\sref{sec-coord-prep}) & Single-server training\\
	\bottomrule[1.2pt]
   \end{tabular}
    \vspace{-1.5em}
\label{tbl-techniques}
\end{table}

\subsection{The \minio cache}
\label{sec-minio}
As datasets increase in size, they cannot be
cached entirely in the memory of a server during training.
In such cases, DNNs  suffer from fetch stalls if the rate of data fetch 
is lower than the rate of compute (despite prefetching and 
pipelining data fetch with compute) as discussed in (\sref{sec-data-stall}). 
Note that, when fetch stalls occur, training is bottlenecked 
by the bandwidth of storage device, therefore it is crucial to 
minimize \vio by maximizing cache hits every epoch.


DNN training frameworks today, rely on the OS Page Cache 
to cache the training dataset. However, 
we tap on the piercing insight of Stonebraker \etal's
pioneering work on database caching~\cite{stonebraker81},
 that \emph{the
abstractions provided by OS can hinder the 
development of efficient databases}, and validate
it in the context of DNN 
workloads. Therefore, we study the DNN data access pattern
to design a domain-specific cache \minio.

\begin{figure}[!t]

  {\includegraphics[width=0.46\textwidth]{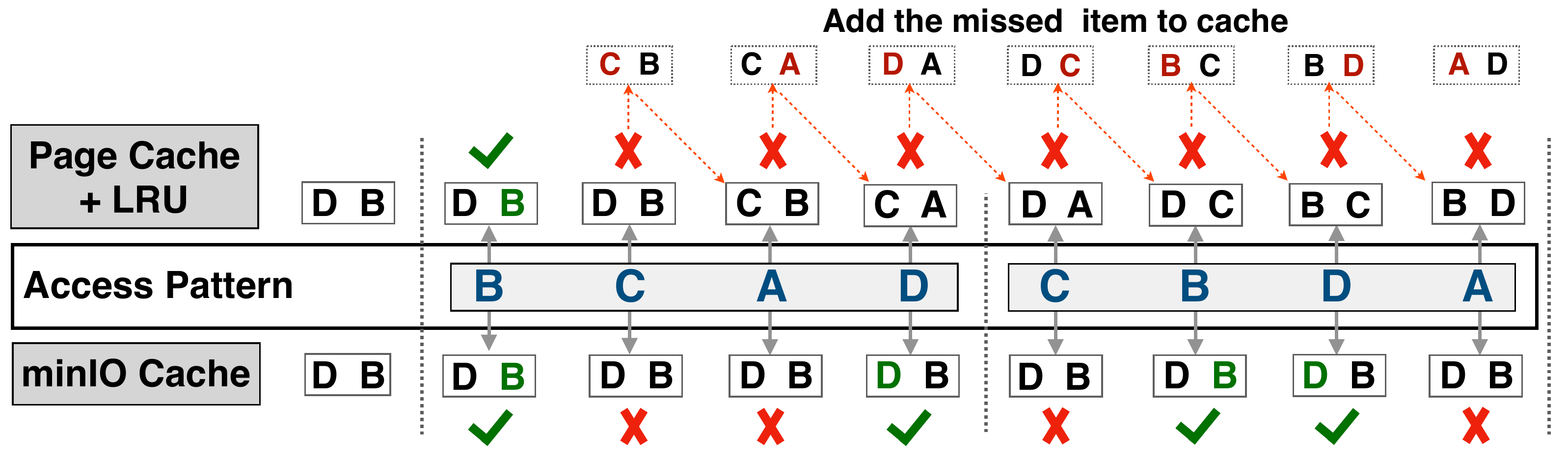}}
   \vspace{-1em}
  \mycaption{Cache hits with \minio}{Cache
  activity for two ``epochs'' of training for page cache and \minio.}
  \label{fig-minio}
  \vspace{-1em}
\end{figure}


OS Page Cache works as follows; whenever a 
data item is read from storage, it  is cached 
in the Page Cache to speed up future accesses. 
When the Page Cache reaches its capacity, a 
\emph{cache replacement policy} decides which of the 
existing items to evict to make space for the new one. 
Linux uses a variant of Least Recently Used (LRU)  
for cache replacement~\cite{linux-lru}.

However, DNN training has a unique data access pattern: it is
\textit{repetitive across epochs} and \textit{random within an epoch}.  Training is split into
epochs: each epoch accesses all the data items in the dataset exactly
once in a random order.
We make a key observation about the DNN access pattern that
is at odds with such OS cache replacement policies. \emph{All} data
items in the dataset have equal probability of access in an
epoch. Therefore, it is \emph{not important} which data item is
cached.  Instead, it is crucial that cached items are not replaced
\emph{before} they are used, to minimize storage \vio per epoch.

Therefore, \minio recommends a simple and unintuitive solution;
\emph{items, once cached, are never replaced} in the DNN cache. \minio
works as follows. In the first epoch of the training job, \minio
caches random data items as they are fetched from storage, to populate
the cache. Once the cache capacity is reached, \minio will not evict
any items in the cache; instead, the requests to other data items default to
storage  accesses. The
items in the \minio cache survive across epochs until the end of
the training job. Every epoch beyond the first gets exactly as many hits
as the number of items in the cache; this reduces the per-epoch disk
\vio to the difference in the size of dataset and the cache.

Figure~\ref{fig-minio} contrasts the caching policy of the OS Page Cache
and \minio. Consider a dataset of size 4 (with items A -- D) and a cache of size
2 (50\% cache). Let's say after warmup, the cache has two items D and
B.  Figure~\ref{fig-minio} shows the state of the cache for two
training epochs.  \minio only incurs capacity misses per epoch (here
2); the Page Cache on the other hand, can result in anywhere between
2-4 misses per epoch because of thrashing. For instance, in the first
epoch, D is in the cache to begin with, but kicked out to make way for
a new item C, and later in the same epoch it is requested again (thrashing).  We
empirically verified this using large datasets and varying cache sizes
(\sref{sec-eval}) and found that Page Cache results in close to 20\%
more misses than \minio due to thrashing.

\minio's no replacement policy simplifies the design of the
cache as we do not need bookkeeping about the time or frequency
of access of data items. 
Moreover, we choose to 
implement \minio in user-space and not as a new replacement policy in the
kernel, making it flexible to use in scenarios where the user has no root privileges to modify the kernel.
The strength of \minio thus lies in its simplicity
and effectiveness.

\subsection{Partitioned \minio Caching}
\label{sec-part-cache}
\minio reduces the amount of disk \vio (fetch stalls) in single-server training. In
distributed training, the dataset is partitioned and processed by a
group of servers. Each server operates on a random shard of the dataset per epoch,
and this partition changes every epoch (~\sref{sec-etl}).
The \minio cache alone, is not efficient in this setting. For example, consider a
distributed training job across two servers, each of which can
cache 50\% of the dataset. In every epoch, each server has to process
a random 50\% partition of the dataset, some of which may be hits in
the local \minio cache but the misses result in storage \vio, which is
expensive and results in fetch stalls.

We observe that the cross-node network bandwidth in publicly available
cloud GPU instances and our clusters(10-40 Gbps) is upto 4\myx higher than the read
bandwidth of local SATA SSDs (530 MBps).  Data transfer over
commodity TCP stack is much faster than fetching a data item from its
local storage, on a cache miss.  Therefore, we can coordinate the remote \minio
caches across all servers.

Partitioned \minio caching works as follows. In the first epoch, the dataset
is sharded across all servers, and each server populates it's local
\minio cache with data items in the shard assigned to it. At the end
of the first epoch, we collectively cache a part of the dataset
of size equal to the sum of capacities of individual \minio caches.
To route data fetch requests to the appropriate server, we
maintain metadata about data items present in each server's cache.
Whenever a local cache miss happens in the subsequent epoch at any
server, the item is first looked up in the metadata; if present,
it is fetched from the respective server over TCP, else from its local storage.
If the aggregate memory on the participating servers is large enough
to cache the entire dataset, then partitioned caching ensures that
there is no storage \vio on any server beyond the first epoch; the entire
dataset is fetched exactly once from disk in the duration of
distributed training. 

%


\subsection{Coordinated Prep}
\label{sec-coord-prep}
Hyperparameter (HP) search for a model involves running several
concurrent training jobs, each with a different value for the HP and
picking the best performing one. Our analysis shows that co-locating HP
search jobs on the same server results in both fetch and prep stalls
(\sref{sec-analysis}) due to lack of coordination in data fetch
and prep among these jobs.

We introduce \emph{coordinated prep} to address this
issue. The idea behing coordinated prep is simple. 
Each job in the HP search operates on the same data;
hence, instead of accessing data independently for each job,
they can be coordinated to fetch and prep the dataset exactly once per
epoch. Each epoch is completed in a synchronized fashion by all HP
jobs; as a result, pre-processed minibatches created by one job can be
reused by all concurrent jobs.

Coordinating HP search jobs must be done carefully to ensure that: 
\emph{each job processes the entire dataset exactly
  once per epoch}.  A naive way of doing this is to pre-process the
dataset once and reuse across all HP jobs and all epochs as 
suggested by prior work~\cite{cerebro, kakaraparthy2019case}. This
approach will not work for two reasons. First, reusing pre-processed
data across epochs may result in lower accuracy, as the random
transformations are crucial for learning. Second, the pre-processed items
are 5--7\myx larger in size when compared to the raw data items.
Caching pre-processed items will overflow the system memory capacity
quickly for large datasets. If we store them on storage, we may incur fetch stalls. 

Coordinated prep addresses these challenges by staging pre-processed
minibatches in memory for a short duration \emph{within an
  epoch}. Since each job has identical per-minibatch processing time,
the minibatch is short lived in the staging area.  Coordinated prep
works as follows.

Each HP search job on a server receives a
random shard of the dataset when they start. Each job fetches and
pre-processes the assigned shard, creating minibatches as they
would normally do. When ready, these minibatches 
are exposed to the other jobs in the cross-job staging area. 
This is  a memory region that is 
accessible to all running jobs on the server.
Additionally, each minibatch has a unique ID and an
associated atomic counter that tracks how many
jobs have used this minibatch so far in the current epoch. 
When a job needs a minibatch for GPU processing, it
retrieves it from the staging area and
updates its usage counter. A minibatch is deleted from the 
staging area when it is used exactly once by all running jobs,
as we want to ensure that it is not used across epochs.
We empirically show in \sref{sec-eval} that the
addition of cross-job staging area does not
introduce additional memory overhead.

Thus, coordinated prep ensures one sweep over the dataset
per epoch for both data fetch and pre-processing, 
eliminating redundant work. Note that coordinated
prep allows addition or removal of jobs only at epoch 
boundaries; this is not an issue because popular
HP search algorithms evaluate the
objective function (for e.g., accuracy),
and make decisions on terminating or
continuing the job at epoch boundaries~\cite{jaderberg2017population,
hyperband}

To handle job failures in HP search, we implement a failure detection module to
monitor the status of running jobs. Every prepared minibatch fetched from the staging area 
has an associated timeout. If any job times out waiting for a minibatch,
it notifies the driver process of a possible failure. 
All the jobs can deterministically identify which job failed to 
populate the batch it is waiting on. If the job indeed failed, a new process is spawned 
to resume data loading for the failed shard.

\subsection{Tying it all together with \sysname}
We implement the three techniques discussed thus far as a part of 
a user-space data loading library, \sysname. We build \sysname
on top of DALI to take advantage of the GPU-accelerated data 
pre-processing operations. 
\sysname can be used as a drop-in replacement
for the default PyTorch dataloader. 

The overall architecture of \sysname is as follows. 
The training dataset resides on a local
storage device like SSD/HDD. If the data resides on a remote
storage service, it is cached in local storage when first
accessed~\cite{blobfuse}. For all later epochs, the data is
fetched from local storage. In each training iteration, a minibatch of
data must be fetched from disk (or cache), pre-processed to apply
random transformations and collated to a tensor that can be copied
over to the GPU for computation. \sysname manages its
own \minio cache of the raw data items (before any stochastic
pre-processing transformations are applied).  The data sampling and
randomization is unmodified; in each epoch, every minibatch is
sampled randomly from the dataset. Every data item is then subjected
to the random pre-processing pipeline specified in the training
workload. The prepared minibatch is then placed in a cross-job staging
area for consumption by the GPU. If a single data-parallel job is
running across multiple GPUs in a server, then the minibatches in the
staging are used exactly once per epoch and discarded; if there are
concurrent HP jobs on a server, then the staging area retains
minibatches until each concurrent job has used it exactly once in the
current epoch. Any minibatch that satisfies this criteria is evicted
from the staging area to make way for newer batches.

	\section{Evaluation}
\label{sec-eval}

We now evaluate the efficacy of \sysname on three different
aspects of the training process:  multi-GPU
training on a single server, distributed training across multiple
servers, and hyperparameter tuning.  We evaluate our techniques on \textbf{nine} models,
performing \textbf{three} different ML tasks (image classification,
object detection and audio classification) on four different
datasets, each over 500GB as shown in Table~\ref{tbl-dataset-models}.
Since DALI strictly outperforms PyTorch DL, we use 
DALI as the baseline in our experiments. 
For each model, we run both CPU-based (all pre-processing on CPU) 
and GPU-based (part of decoding and all other transformations on GPU)
 mode of DALI, and present the best of the two results.

\vheading{Experimental setup}.  We evaluate \sysname on two
representative server configurations from Microsoft clusters
 (Tbl~\ref{tbl-analysis-sku}) each with 500 \gb DRAM, 24 CPU
cores, 40 Gbps Ethernet, eight GPUs, and 1.8 \tb of storage
space. \confssd uses V100 GPUs and a SATA SSD, while \confhdd uses
1080Ti GPUs and a magnetic hard drive. 
We use the same training methodology we used for analysis (\sref{sec-methodology}).
We seek to answer the following questions:

\vspace{-1em}
\setlength\mylength{\dimexpr.5\columnwidth-10\tabcolsep-0.5\arrayrulewidth\relax}
\newcolumntype{M}[1]{>{\centering\arraybackslash}m{#1}}
\begin{table}[!h]
    \small
 	\centering
 	\ra{0.95}
 	 \begin{tabular}{!{\VRule[1pt]}m{0.42\textwidth}!{\VRule[1pt]}m{0.03\textwidth}!{\VRule[1pt]}}
 	\specialrule{1.2pt}{0pt}{0pt}
 	 \rowcolor{white}
 	
 		How does the \minio cache affect multi-GPU training on a
        server? & \sref{sec-mint} \\
         \specialrule{0.5pt}{0pt}{0pt}
 		How does partitioned caching improve training time for jobs
 		distributed across multiple servers? & \sref{sec-dist-mint}\\
 		 \specialrule{0.5pt}{0pt}{0pt}
 	   How does coordinated prep benefit HP search?
 		& \sref{sec-unified} \\
 		 \specialrule{0.5pt}{0pt}{0pt}
 	   Does \sysname affect DNN training accuracy?
 		& \sref{sec-eval-accuracy}\\
 		 \specialrule{0.5pt}{0pt}{0pt}
 		Does \sysname enable better resource utilization compared to
 		DALI? & \sref{sec-resource}\\
 		 \specialrule{0.5pt}{0pt}{0pt}
 		Does \sysname accelerate training on 
 		ML servers
 		like the DGX-2? 
 		& \sref{sec-dgx2}\\
 		 \specialrule{1.2pt}{0pt}{0pt}
 	\end{tabular}
\label{tbl-eval-q}
\vspace{-1em}
\end{table}

 \begin{figure*}[!t]

  \centering 

  \includegraphics[width=.9\linewidth]{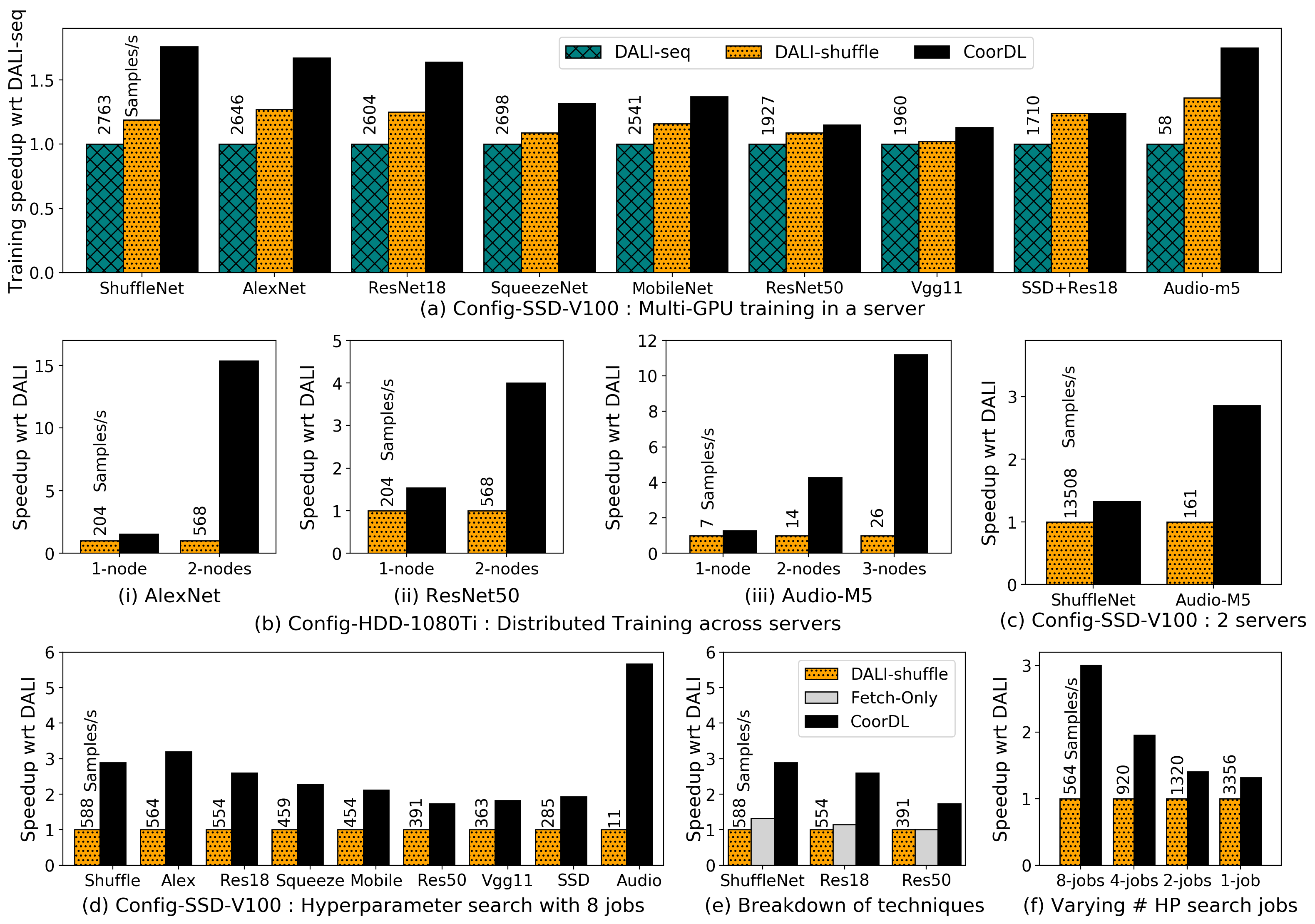} 
   
 \vspace*{-1em}    
 \caption{This graph compares DALI against
      \sysname for a variety of training scenarios; single server, multi-server
      and HP search, across 2 
      clusters and 9 models. \sysname significantly accelerates training 
      by eliminating redundant data fetch and prep.
  }
  \label{fig-eval}

\end{figure*}
\begin{figure*}[!htb]
\minipage[b]{0.33\textwidth}
  \includegraphics[width=\linewidth]{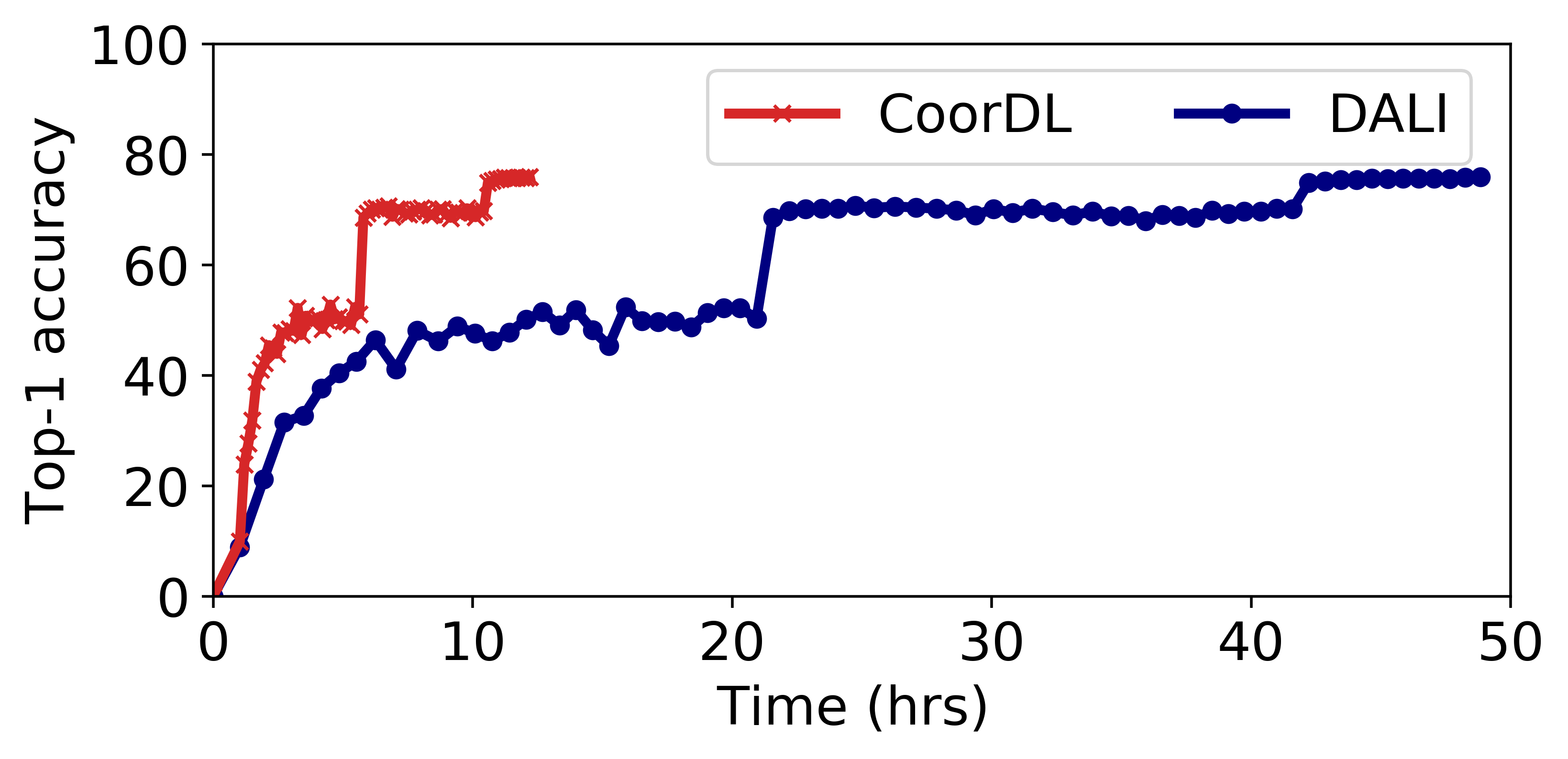}
  \vspace*{-2.5em}
  \caption{Training to accuracy}\label{fig-accuracy}
\endminipage\hfill
\minipage[b]{0.33\textwidth}
  \includegraphics[width=\linewidth]{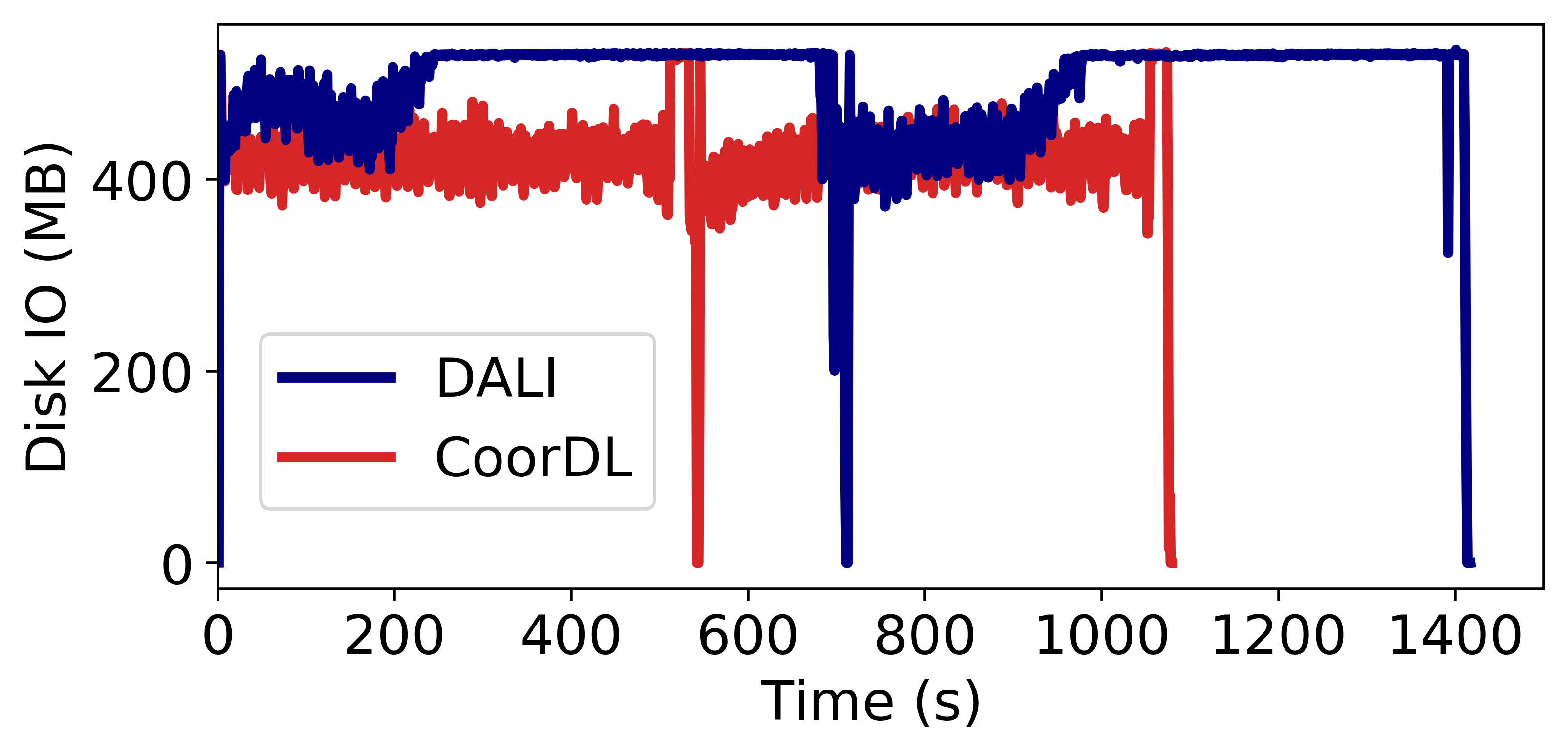}
   \vspace*{-2.5em}
  \caption{Disk \vio profile}\label{fig-mint-io}
\endminipage\hfill
\minipage[b]{0.33\textwidth}%
  \includegraphics[width=\linewidth]{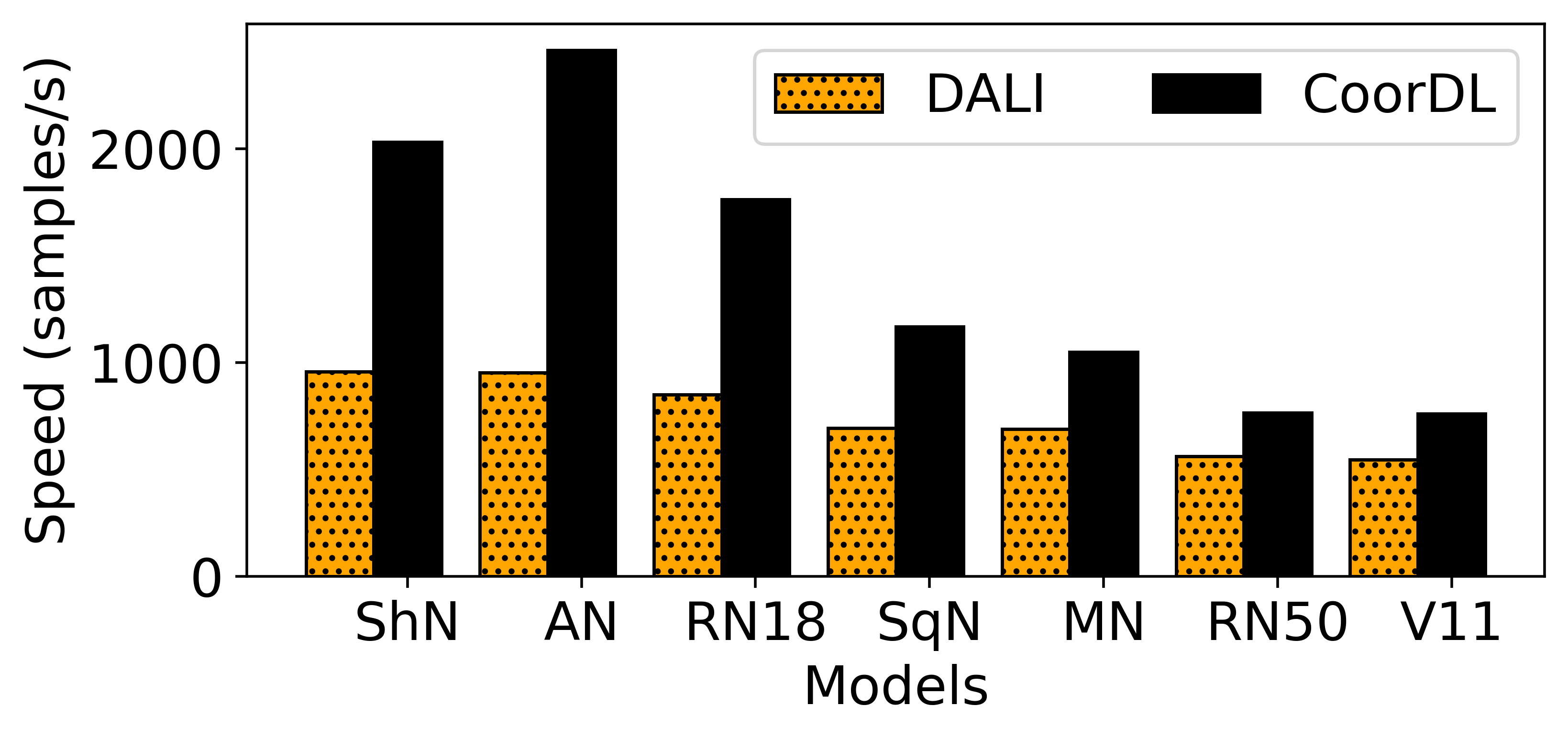}
   \vspace*{-2.5em}
  \caption{Evaluation on DGX-2}\label{fig-dgx2}
\endminipage
   \vspace*{-1em}
\end{figure*}

\subsection{Single-server Multi-GPU training}
\label{sec-mint}

\sysname speeds up a single-server training job by reducing cache
misses using the \minio cache. Figure~\ref{fig-eval} (a) plots the
relative speedup with respect to DALI while training the image
classification and object detection models on the OpenImages dataset,
and audio classification on FMA dataset. We evaluate \minio against
two modes of DALI. DALI's default mode is \emph{DALI-seq}, where it
reads data sequentially off storage and shuffles them in
memory~\cite{dali-shuffle}. \emph{DALI-shuffle} accesses the dataset
in a randomized order (doing random reads, similar to the native dataloader of PyTorch).

\minio results in upto 1.8\myx higher training speed
compared to DALI-seq by eliminating thrashing on \confssd. When the
image classification models are trained with ImageNet-22k dataset,
\sysname results in up to 1.5\myx speedup. On \confhdd, \sysname
accelerates ResNet50 training on 
OpenImages by 2.1\myx compared to DALI-seq and 1.53\myx 
compared to DALI-shuffle respectively.

 
\vheading{Reduction in cache misses}. We measure the disk \vio and
number of cache misses when training ShuffleNet on OpenImages dataset
on \confssd. This server can cache 65\% of the dataset. \sysname
reduces misses to the minimum number of 35\%, resulting in 225 GB of
\vio. In contrast, DALI-Seq results in 66\% cache misses, increasing
\vio by 87\% to 422 GB; DALI-shuffle results in 53\% cache misses,
increasing \vio by 50\% compared to \sysname to 340 GB.

Note that, when the whole dataset does not fit in memory, DALI-shuffle
performs better than DALI-seq (because sequential access is a
pathological case for the Linux LRU page cache). Therefore, our
evaluation in the rest of this section compares \sysname to the
stronger baseline, DALI-shuffle.

\subsection{Multi-Server Distributed Training}
\label{sec-dist-mint}
We now evaluate \sysname on a distributed training scenario.
The lack of cache co-ordination between the participating servers
results in fetch misses that lead to disk I/O.  \sysname uses
partitioned caching to avoid redundant \vio.


Figure~\ref{fig-eval}(b) shows that \sysname improves the throughput
of distributed training jobs by upto 15\myx (AlexNet on OpenImages)
when trained across two \confhdd servers (16 GPUs). On \confhdd
servers, 65\% of the OpenImages dataset can be cached on a single
server; and it can be fully cached in the aggregated memory of two
servers. Therefore, \sysname moves the training job from being I/O
bound to GPU bound.

When trained across two servers on \confssd, \sysname accelerates
ShuffleNet on ImageNet-22k by 1.3\myx, and Audio-M5 on FMA by 2.9\myx.
The relative gains are lower on \confssd because the cost of a fetch
miss is lower on SSDs due to its high random read throughput, as
compared to HDDs on \confhdd.

\subsection{Hyperparameter Search}
\label{sec-unified}

Figure~\ref{fig-eval} (d) plots the relative increase in throughput of
individual jobs across several models when eight concurrent HP search
jobs are run on a \confssd server. On less computationally complex
models like AlexNet and ShuffleNet, \sysname increases training speed
by 3\myx, because these models are originally CPU bound due to prep.
For the audio model, \sysname increases the training speed by
5.6\myx. \sysname reduced the total disk \vio from 3.5TB to 550GB. 
Similarly, on \confhdd, \sysname results in 5.3\myx faster training on
the audio model, and 4.5\myx faster training on ResNet50.
On \confhdd, \sysname results in 5.3\myx
faster training on the audio model, and 4.5\myx faster training on
ResNet50 by coordinating data fetch and prep.

\vheading{Split of coordinated prep benefits}.
Next, we show the breakdown of speedup due to coordination
of data fetch and prep during HP search.
When fetch is coordinated, concurrent jobs use data fetched by other jobs; but each job performs its own data prep. 
\sysname coordinates both; eliminating redundant fetch and prep.
Figure~\ref{fig-eval} (e) plots the results on \confssd. 
In this case, data stall is dominated by 
prep, which \sysname mitigates 
unlike prior work like Quiver~\cite{kumar2020quiver} that only coordinates fetch.

\vheading{Multi-GPU HP search jobs}.  Figure~\ref{fig-eval} (f)
shows training with different configs of HP search jobs on a machine; 8 1-GPU jobs, 4 2-GPU jobs, 2
4-GPU jobs, or 1 8-GPU job for AlexNet on OpenImages. For a single job
case, the benefit is due to the \minio cache; in other configs,
it is due to coordinated prep. More the number of concurrent jobs, higher the benefit
with coordination.

\subsection{Training to Accuracy with \sysname}
\label{sec-eval-accuracy}

\sysname does not change the randomness of data pre-processing
 involved. Its techniques do not affect the learning algorithm.
To demonstrate this, we train ResNet50 to accuracy on ImageNet-1K
using 16 GPUs across two \confhdd servers, where each server is
capable of caching 50\% of the dataset.
Figure~\ref{fig-accuracy} shows that \sysname reduces the time to
target accuracy (75.9\%) from two days to just 12 hours (4\myx
better), due to partitioned caching.

\subsection {Resource Utilization}
\label{sec-resource}

\vheading{\minio results in lower disk I/O and better CPU utilization}.
Figure~\ref{fig-mint-io} shows the I/O for two epochs of training
ResNet18 on OpenImages on \confssd.  The I/O behavior is similar
across models and server configurations.
DALI observes cache hits at the beginning of the epoch, but
soon becomes I/O bound (disk bandwidth: 530 MB/s).
Since \minio is caching a random subset of the dataset, 
cache hits are uniformly distributed across the epoch in
\sysname. This results in a predictable I/O access pattern and faster
training (epochs end earlier in Figure~\ref{fig-mint-io}).
Profiling the CPU during training shows that the pre-processing
threads in DALI are often stalled waiting for I/O. Since \minio
reduces the total disk I/O, \sysname is able to better utilize the CPU
for pre-processing. The combination of lower disk I/O and
better CPU utilization leads to shorter training times when using \sysname.




\vheading{\sysname uses a fraction of available network
  bandwidth}. \sysname shards the dataset equally among all servers in
distributed training to ensure load balancing. We track the network
activity during the distributed training for ResNet50 on OpenImages
across two, three, and four servers with DALI and \sysname. \sysname
used 5.7 Gbps per server of network bandwidth (14\% of the 40 Gbps
available). DALI used 1.18 Gbps  of network
bandwidth per server. \sysname used 4.8\myx higher network bandwidth to train
4.3\myx faster.





\vheading{Co-ordinated prep has low memory overhead}.  By design,
co-ordinated prep has the same memory requirements as DALI. To
experimentally validate this, we track the memory utilization of
running hyperparameter search on AlexNet on OpenImages on a \confssd
server using eight concurrent jobs.
\sysname uses 5 GB of extra process memory to store prepared
mini-batches in memory until all hyperparameter jobs consume it. We
reduce the cache space given to \sysname by 5 GB (keeping the total
memory consumption same for \sysname and DALI). Despite the lower
cache space, \sysname still accelerated training by 2.9\myx.





\subsection{\sysname on DGX-2}
\label{sec-dgx2}
We now compare \sysname against DALI on the bleeding-edge
ML optimized server DGX-2 while performing HP search 
across 16 GPUs using OpenImages dataset. 
Since this dataset can be fully cached in the memory of
DGX-2 (1.5TB DRAM), we observe no stalls due to data fetch beyond
the first epoch. However, the imbalance in the ratio of CPU-GPU 
results in prep stalls which \sysname mitigates by coordinating
pre-processing.  
\sysname accelerates HP search by 1.5\myx -- 2.5\myx over DALI
by eliminating redundant data prep, enabling efficient usage of CPU
to mask prep stalls. 

	\section{\revision{Discussion}}
\label{sec-disc}
\revision{
Our analysis of data stalls reveals several key insights to
utilize the computational capabilities of GPUs by minimizing
data stalls. While we explore ways of mitigating
data stalls in a user-space library \sysname, we believe there is
more to be done. }

\vheading{\revision{Decoded cache to reduce pre-processing overhead}}. 
\revision{\sysname does not address the high cost of decoding/decompressing
raw images, which our analysis identifies to be the most expensive
operation during data prep. A future direction is to evaluate the 
benefits of caching decoded 
data items instead of the current approach of caching raw encoded items. 
Since decoding is deterministic, it is possible to
cache it across epochs. However, this is non trivial; 
decoding increases the dataset size by  5 -- 7 \myx.
It is an interesting future direction to enable decoded caching
without incurring the high space overhead, possibly using serialized
data formats.}

\vheading{\revision{Automatic prep offload to GPUs}}. 
\revision{Data pipelining frameworks like DALI have the ability to perform
certain image and audio based pre-processing (prep) such as crop, flip, and other transformations 
on GPU accelerators. However, there is a memory-performance tradeoff in deciding how many
steps of prep are offloaded to the GPU for two reasons. (1) Performing prep
at the GPU takes up a part of the already scarce GPU memory which may result in training with
lower batch sizes, thereby affecting training efficiency. (2) Prep at the GPU
may interfere with the computations performed by the learning algorithm; this adversely affects the 
overall throughput of training for computationally expensive and deeper networks. Therefore, the split of 
prep operations must be carefully chosen  considering the model's architecture, batch size, 
and data stalls. While this split is determined manually by trial-and-error today, automating it with a careful eye on GPU and CPU utilization is an interesting future direction.}

\vheading{\revision{Minibatch as a service}}. 
\revision{Our analysis shows that the imbalance in CPU cores per GPU in ML optimized servers result
in data stalls for several models. In such cases where single-host capacities are maxed out, a viable approach
is to offload data prep to other idle host machines in a cluster. This is especially useful in 
production clusters with high-bandwidth ethernet, where several jobs use the same dataset and similar pre-processing pipelines; a dedicated set of servers can be used to centrally pre-process minibatches of data, while the training jobs can request minibatch as a service, thereby
entirely disaggregating learning from data management.}

\vheading{\revision{Cost-peformance tradeoff of upgrading hardware}}. 
\revision{Our analysis finds that data stalls squander away the improved performance
of faster, expensive GPUs, resulting in lower value/\$ spent as shown in 
Figure~\ref{fig-analysis-prep-res18} (b). Therefore, it may be economical to train some models
on slower, less expensive GPUs with no data stalls, rather on underutilizing 
the accelerator capabilities due to stalls on faster, expensive GPUs. In practice, such techniques
may improve the overall efficiency in multi-tenant clusters by assigning jobs to accelerators in such a 
way that they maximize GPU utilization.}

\vheading{\revision{Data stalls in inference}}. 
\revision{This work addresses data stalls in the training pipeline which have three distinct features
from inference. (1) Training requires a large volume of data samples, (2) performs a larger set of data prep 
for every batch, and (3) requires backpropagation during the learning phase. While inference jobs require fewer prep steps
per sample or batch, it also performs lesser GPU computation compared to training. Moreover, the limited memory and compute availability at edge devices may also introduce data stalls in inference. We hope our analysis encourages similar
research and possibly unique optimizations in inference land.}

\vheading{\revision{Trade-off between convergence rate and epoch time for other SGD variants}}. 
\revision{This work focuses on the most common case of mini-batch SGD with a random shuffling of the data in every epoch which is the default for the models we analyzed. A future direction is to understand the impact of relaxing the ETL requirements assumed in this work (such as random prep and shuffling every epoch) on epoch time and model convergence. Although relaxing these constraints may reduce data stalls and hence epoch time, it may prolong convergence, or affect the accuracy of some models. It is worth investigating this behavior theoretically and empirically.}

	\section{Related Work}
\label{sec-related}
To the best of our knowledge, this paper presents the first comprehensive
analysis of data stalls in DNN training. We place our work in the
context of prior work.


\vheading{Optimizing remote storage via local
  caching}. Quiver~\cite{kumar2020quiver} uses local SSD caches to
eliminate the impact of slow reads from remote storage. The best case
for Quiver is when the dataset is completely cached on local storage;
our system starts from this baseline and further improves
performance. Quiver does not consider or optimize prep stalls, only
handling fetch stalls. 


\vheading{Partitioned caching}.  Cerebro~\cite{cerebro} introduces a
new parallel SGD strategy for model selection tasks. It partitions the
dataset across the servers in a cluster and hops the models from one
server to other, instead of shuffling
data. 
Cerebro does not improve performance for DNN training on a single
server, while \sysname optimizes performance in this scenario.
Furthermore, Cerebro is designed for a specifc scenario - distributed
model search; on the contrary, our analysis and \sysname have a
broader scope.  DeepIO~\cite{zhu2018entropy} uses a partitioned
caching technique for distributed training with remote data, but
relies on specialized hardware like RDMA. In contrast, our work shows
that it is possible to mask fetch stalls using commodity TCP
stacks. 




\vheading{Redundancy in DNN training}. 
Prior work like Model Batching~\cite{narayanan2018accelerating} 
addressed redundancy in model search 
when running multiple DNNs together on a single GPU, by
sharing GPU computation across jobs. Our analysis looks at 
the setting where GPUs are not shared between jobs.
OneAccess~\cite{kakaraparthy2019case}  is a preliminary study 
that makes a strong case for 
storing pre-processed data across epochs to reduce prep stalls; however such an approach precludes commonly used online data pre-processing techniques 
 and this can affect model convergence. 
In contrast, \sysname carefully
eliminates redundancy while preserving accuracy and providing significant
speedups.

\vheading{Hardware solutions to fetch stalls}. New hardware like NVIDIA's
Magnum IO~\cite{magnum}, and PureStorage's AIRI~\cite{airi}
provide high throughput storage solutions to address  fetch stalls. While these fast hardware may mask 
fetch stalls in some models, they may not help if the model is bottlenecked on prep stalls. 
Our work mitigates data stalls  with existing 
commodity servers as opposed to relying on expensive hardware solutions, by efficiently using available hardware.

\vheading{Optimizing DNN training time}.  
 A number of solutions have been proposed to reduce the training time
 for DNNs including specialized hardware~\cite{jouppi2017datacenter,
   jeon2019analysis, tpu, ovtcharov2015accelerating, nurvitadhi2017can,
   putnam2014reconfigurable, cerebras, graphcore}, parallel
 training~\cite{krizhevsky2012imagenet, dean2012deepbelief,
   chilimbi2014adam, jia2018beyond, krizhevsky2014one,
   narayanan2019pipedream, huang2018gpipe}, GPU memory
 optimizations~\cite{vdnn,chen2016training,gist2018}, lowering
 communication overhead~\cite{lin2017dgc, PoseidonATC2017,
   hashemi2018tictac, jayarajan2019priority, nccl, wangt2018blink}, 
, and operator optimizations~\cite{vasilache2018tensor, chen2018tvm,
   jia2019taso}. This paper presents a new point in this spectrum, \textit{data stalls}.


\vheading{Domain specific caching}. 
The idea of designing a caching policy that
is aware of application semantics is not new. Stonebraker \etal
highlighted the importance of domain-aware caching for databases~\cite{stonebraker81}. Tomkins \etal show that
informed prefetching and caching  in file systems can  reduce the
execution time of \vio-intensive
applications~\cite{tomkins1997informed}. 
Our work draws parallels to such techniques by first
understanding DNN access pattern and then devising a
caching policy based on these observations.

\section{Conclusion}
 We present the first detailed study of data
stalls in several DNNs, and show that it
accounts for up to 65\% of the training time. The insights from
our study, guide the design of a coordinated caching and
pre-processing library, \sysname, that can
accelerate DNN training by mitigating data stalls. 
\sysname accelerates
 training by up to 15\myx for distributed training
across two servers, and 5.3\myx for HP search (on the audio
model), by coordinating data fetch and prep
across jobs. 
The techniques behind \sysname are
simple and intuitive, easing adoption into production systems.

	\section*{Acknowledgements}

This work was done during an internship at Microsoft Research as part of Project Fiddle. We thank all the anonymous VLDB reviewers,  members of the UT SaSLab, Jorgen Thelin, Jack Kosian, Deepak Narayanan, Keshav Santhanam and many of our MSR colleagues for their invaluable feedback that made this work better. We sincerely thank MSR Labs for their generous, unwavering support in procuring the many resources required for this work.
	
	\balance
	\bibliographystyle{ACM-Reference-Format}
	\bibliography{all}
	\newpage
	\ifthenelse{\boolean{addappendix}}{
\section*{Appendix}
\label{sec-appendix}
\appendix

This document contains supplementary material, describing experiments that were omitted
in the paper for brevity. 
\section{Analysis of Data Stalls}

Our paper shows the analysis of data stalls 
in DNN training across various models, datasets,
and hardware configurations. Here, we provide
additional analysis of prep stalls such as increasing the
number of CPU cores per GPU beyond 3, 
and the data pipeline rates for different models, datasets and number
of GPUs.

\subsection{Data pipeline rates}
\label{sec-app-rates}
Using \tool, we measure the rate of different components
in the data pipeline as shown in Figure~\ref{fig-rate}. 
$G$ indicates the GPU rate, $P_g$ represents the pre-processing rate
when using the GPU-mode of DALI, $P_c$ is the prep rate for the CPU-mode of 
DALI, and $F$ is the effective fetch rate.
We show these
quantities on \confssd for 7 different models and varying GPU count for the 
OpenImages dataset in Figure~\ref{fig-rate-oi}, ImageNet-22k in
Figure~\ref{fig-rate-img-22k}  and ImageNet-1K in Figure~\ref{fig-rate-img1k}.
The configurations and cache size for each dataset is shown in Table~\ref{tbl-config-rates}
When the GPU count is reduced, all other resources (CPU, and memory)
are also proportionately reduced to maintain the SKU.

$P = max(P_g, P_c)$, the best of CPU or GPU-based prep

The speed of the data pipeline is given by : $min(P, F)$

Data stall exists in a given configuration if $G > min(P, F)$

We see that across several models and configurations, data stalls are prominent.
The fetch rate shown here, assumes an efficient in-memory cache like
\minio that provides x hits per epoch if the cache has x items. 
If we rely on OS Page Cache, fetch rates observed are ~20\% lower, resulting in 
higher fetch stalls.

As we decrease the number of GPUs, the ingestion rate at the GPU $G$ decreases,
as they process data slower due to decrease in parallelism. In all configurations,
the bandwidth of the storage device remains constant; therefore as we reduce the
number of GPUs, $F$, catches up with $G$, and the bottleneck in the training pipeline
shifts to pre-processing.

There following are the key takeaways from the rate graphs.
\begin{itemize}
	\item Several models experience data stalls due to fetch or prep across a range of GPU configurations and datasets.
	\item Rich datasets like OpenImages ( higher per-image size) result in higher data stalls as shown in 
	Figure~\ref{fig-rate-oi}
	\item Even computationally expensive models like ResNet50 and VGG experience data stalls due to 
	the costly pre-processing, despite using state-of-the-art data pipelines.
	\item GPU based pre-processing hurts the performance of some models like ResNet50 and VGG due to
	intereference with GPU computation.
\end{itemize}

\subsection{Training on servers with high CPU count}
Typically, servers optimized for ML training (for e.g., NVIDIA
DGX-2) have 3 CPU cores per GPU~\cite{dgx2}. However,
some cloud providers like AWS have servers with
8 GPUs and 32 CPU cores (64 vCPUs), that results in 
4 cores (or 8 vCPUs) per GPU.
We analyze prep stalls in one such server with 
8 V100 GPUs, 64 vCPUs, and 500\gb DRAM.

\vspace{3em}

\begin{figure}[!h]
  \centering 
   \includegraphics[width=.48\textwidth]{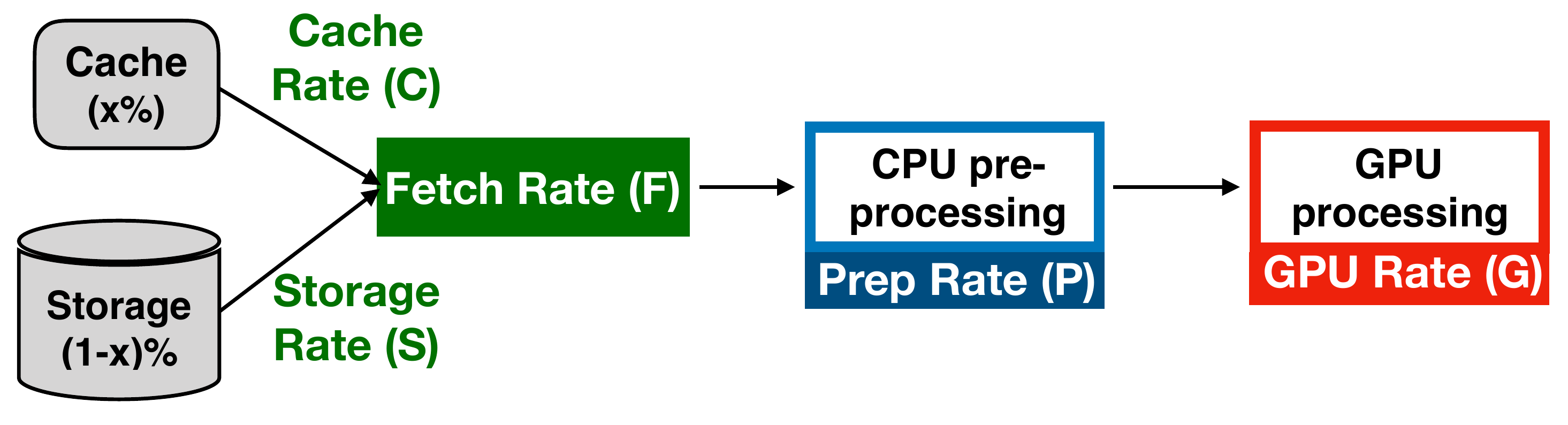}
  
  \mycaption{Data Pipeline in DNN training}{This figure shows the different hardware components 
  involved in DNN training and the throughput of each component.}
  \vspace{-1mm}
  \label{fig-rate}

\end{figure}

\newcolumntype{P}[1]{>{\centering\arraybackslash}p{#1}}
\begin{table}[!h]
  \small
  \centering
  \ra{1.3}
   \caption{The table shows the number of GPUs, CPUs, and the percentage of dataset cached in memory for different configurations on \confssd.}
   \vspace{-1em}
  \begin{tabular}{@{}cccccc@{}}
	\toprule[1.2pt]
        Num & Num & Mem & \multicolumn{3}{c}{\% dataset cached} \\
           GPU & CPU& (GB) & OpenImages & ImageNet-22K & ImageNet-1K \\
          \midrule
          8 & 24 & 500 & 65\% & 35\% & 100\% \\
          4 & 12 & 250 & 32\% &  17\%  & 100\% \\
          2 & 6 & 125 & 16\%  &  9\% & 70\% \\
          1 & 3 & 62.5 & 8\% &  5\% & 35\% \\
  \bottomrule
   \end{tabular}

\label{tbl-config-rates}
\end{table}

\begin{figure*}[!t]
  \centering 
   \includegraphics[width=.93\textwidth]{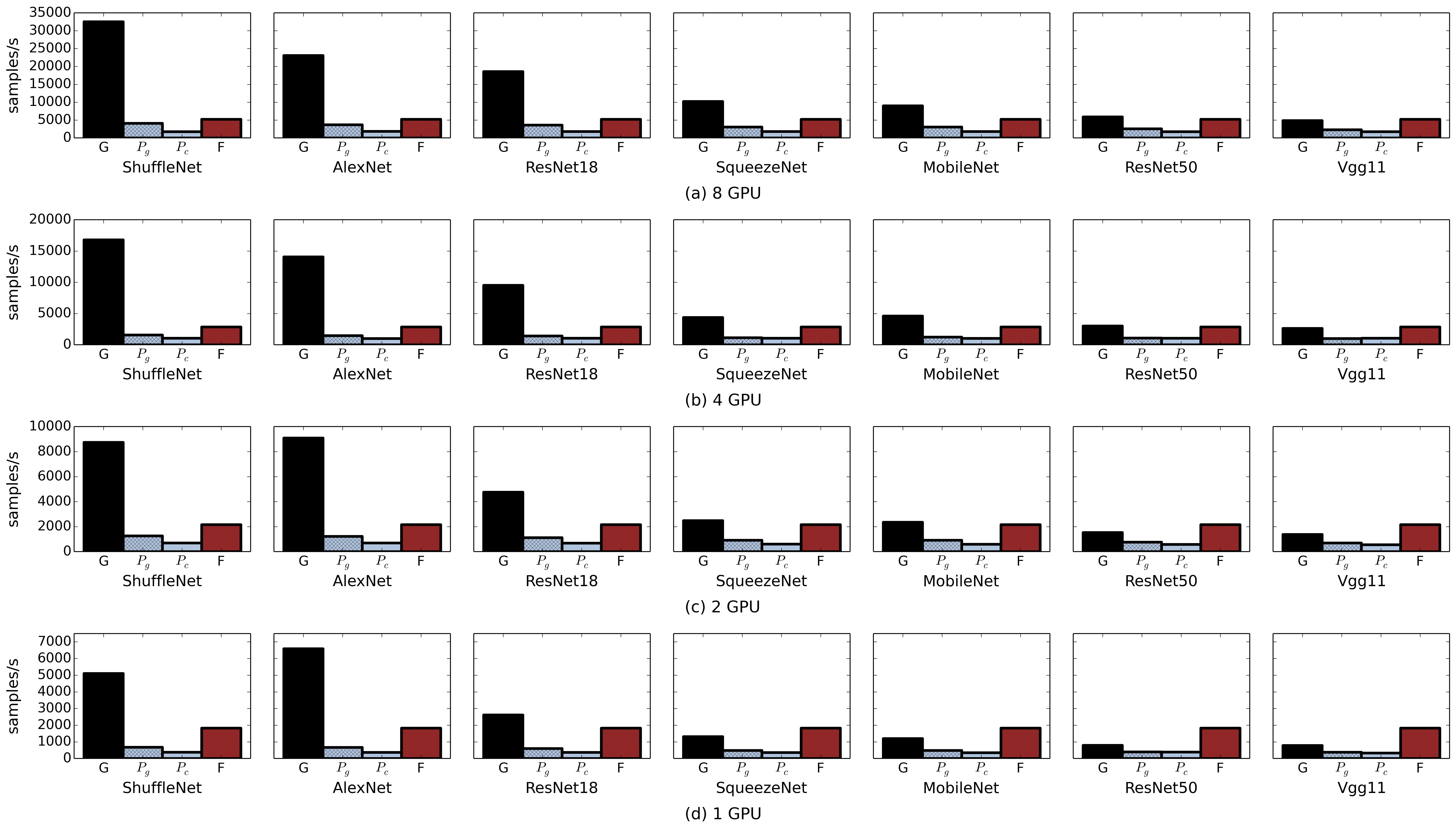}
    \vspace{-1em}
  \mycaption{Data pipeline rates with OpenImages}{This graph shows that several models across various configurations have data stalls; i.e., the black bar ($G$) is greater than the minimum of blue($P$) and red($F$) bars.}

  \label{fig-rate-oi}

\end{figure*}
\begin{figure*}[!t]
  \centering 
   \includegraphics[width=.92\textwidth]{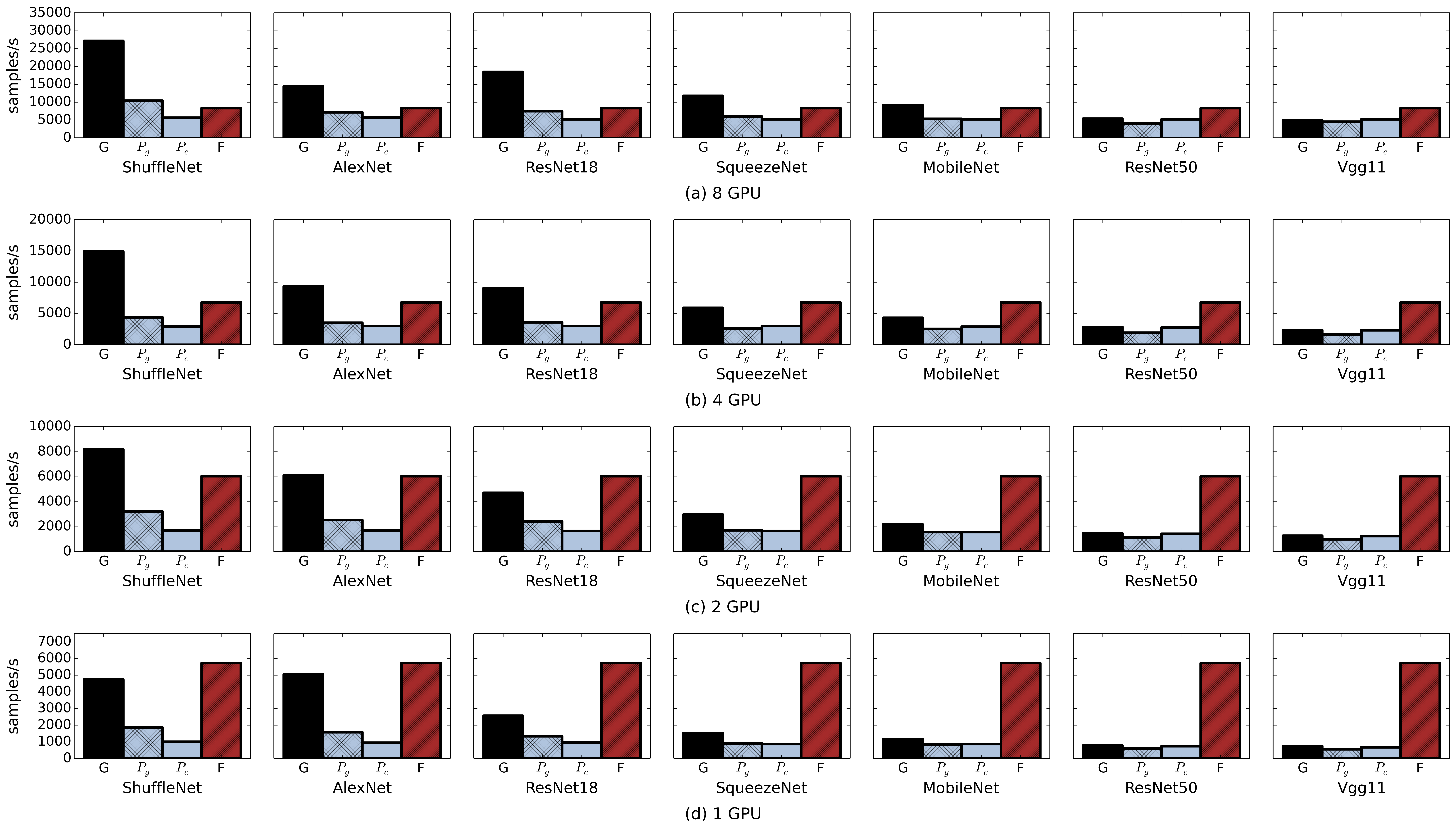}
    \vspace{-1em}
  \mycaption{Data pipeline rates with ImageNet-22K}{This graph shows that several models across various configurations have data stalls due to fetch and prep; i.e., the black bar ($G$) is greater than the minimum of blue($P$) and red($F$) bars.}

  \label{fig-rate-img-22k}

\end{figure*}
\begin{figure*}[!t]
  \centering 
   \includegraphics[width=.93\textwidth]{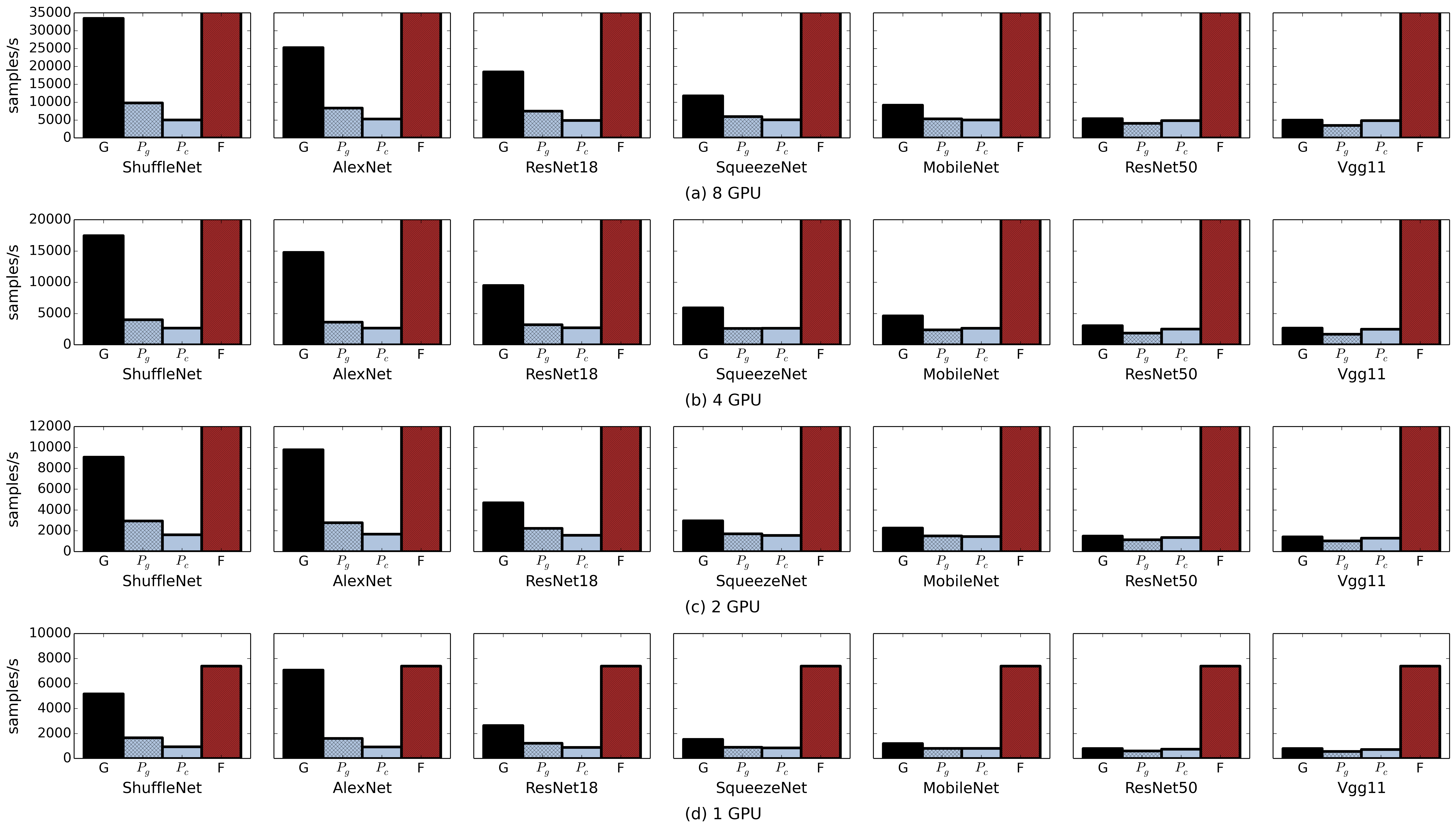}
  \vspace{-1em}
  \mycaption{Data pipeline rates with ImageNet-1K}{This graph shows that several models across various configurations have prep stalls; i.e., the black bar ($G$) is greater than the  blue($P$) bars. Since the dataset fits in memory in most configurations, $F$ is high. }
  \vspace{-1em}
  \label{fig-rate-img1k}

\end{figure*}

\begin{figure}[!t]
  \centering 
   \includegraphics[width=.47\textwidth]{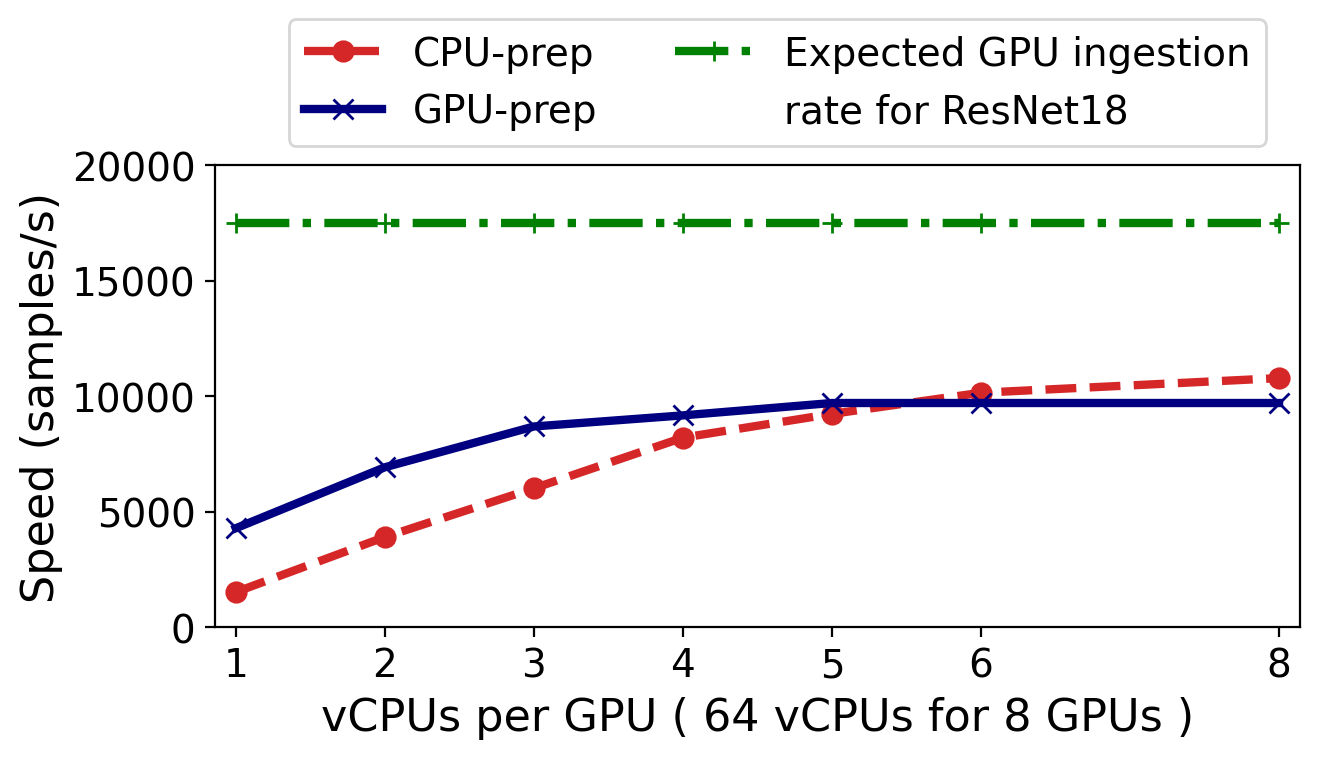}
  
  \mycaption{Impact of CPU on prep}{This graph plots the epoch time for ResNet18 on \confssd as we vary the number of vCPUs per GPU. Although epoch time decreases with higher vCPUs, at 8vCPUs per GPU, ResNet18 has 37\% prep stalls. With the GPU-prep of DALI, we do not increase threads beyond 5 per GPU as it results in GPU OOM.}
  \vspace{-1mm}
  \label{fig-analysis-prep-threads}

\end{figure}
\begin{figure*}[!t]
  \centering 
   \includegraphics[width=.85\textwidth]{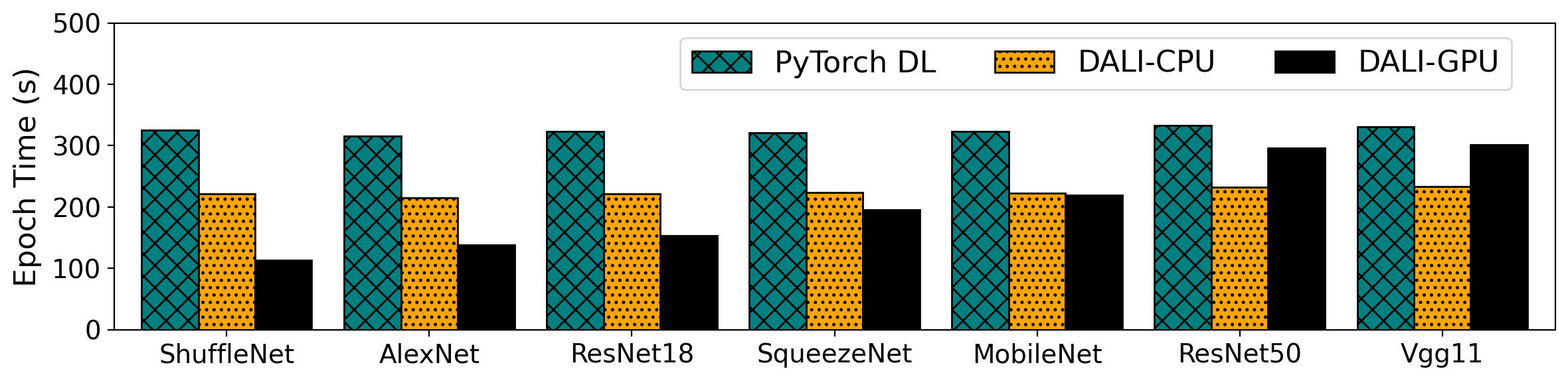}
  
  \mycaption{Epoch time with PyTorch and DALI}{This graph plots the epoch time for various image classification models with native PyTorch DL and
  DALI (CPU-prep and GPU-prep) on \confssd. DALI provides significant speedup over PyTorch even in its CPU mode due to the optimized nvJPEG decoding library. For compute heavy models like ResNet50, GPU based pipeline hurts performance because there is no idle time at the GPU that can be used for pre-processing, and thus interferes 
  with GPU computation.}
  \vspace{-1mm}
  \label{fig-analysis-pytorch}

\end{figure*}

Figure~\ref{fig-analysis-prep-threads} shows the training speed
for a Resnet18 training job as we vary the number of 
vCPUs per GPU for both CPU-based and GPU-based prep pipeline 
with DALI. Note that the dataset is fully cached in memory and
there are no fetch stalls in this experiment. 

Resnet18 has 50\% prep stall for 3 CPU cores
per GPU, when GPU-based prep is used
(shown by the GPU ingestion rate in Figure~\ref{fig-analysis-prep-threads} ). 
With 8 vCPUs per GPU, prep stalls reduced to 37\%, but did not vanish.
Note that, pre-processing with CPU scales linearly upto the
number of cores (here 4 per GPU), beyond that, 
hyperthreading does not result in linear gain in performance.
Increasing the number of pre-processing threads in the server
from 32 to 64 increased pre-processing speed only by 30\%.
In this experiment, we did not increase pre-processing threads 
per GPU beyond 5 in the GPU-prep mode of DALI, as it resulted in
higher GPU memory consumption for prep and hence OOM.
All the experiments presented in our main submission used
3 physical CPU cores per GPU (with GPU-prep of DALI where beneficial). 
This is only 25\% slower than using 8 vCPUs per GPU (as shown
in Figure~\ref{fig-analysis-prep-threads}).
Additionally, the prep stall shown here is for 
ImageNet-1K; with richer datasets like OpenImages 
(higher per-image size), prep stalls increase further.

\subsection{Comparing PyTorch DL with DALI}
PyTorch has two different native modes for data parallel training; DataParallel (DP) and DistributedDataParallel (DDP). DP is usually slower than DDP even on a single server due to
the Global Interpreter Lock (GIL) contention across threads, 
and additional overhead introduced by scattering inputs and gathering outputs across GPUs~\cite{ddp-dp}.
Figure~\ref{fig-analysis-pytorch} shows the 
epoch time for 7 different image classification models using the ImageNet-1K dataset(fully cached in memory) using native PyTorch DL with the faster DDP mode and
DALI. PyTorch DL uses the Pillow library~\cite{clark2015pillow} and TorchVision~\cite{torchvision} for
image decoding and pre-processing while 
DALI uses the optimized nvJPEG library~\cite{nvjpeg}, therefore
resulting in faster pre-processing even when using only CPU. When the GPU based DALI 
pipeline is used, training time further drops 
due to reduction in prep stalls. However, note that there are two downsides to using DALI's
GPU based prep. First, it takes up 2-5GB of
additional GPU memory for pre-processing, 
the luxury of which may not be available for all models and GPUs, as GPU memory is limited.
Second, scheduling pre-processing on the GPU hurts models like ResNet50,
as they are already heavy on GPU computation. 
In all our analysis and evaluation presented in the paper, we run with both GPU and CPU based 
DALI pipeline and present the best of the two results (CPU-based prep was
faster on ResNet-50 and VGG11).

\section{Evaluation of \sysname against DALI}
Our paper evaluates \sysname against DALI in various
training scenarios. This document provides a more
detailed evaluation of some aspects of \sysname.

\begin{figure}[!t]
  \centering
  {\includegraphics[width=0.5\textwidth]{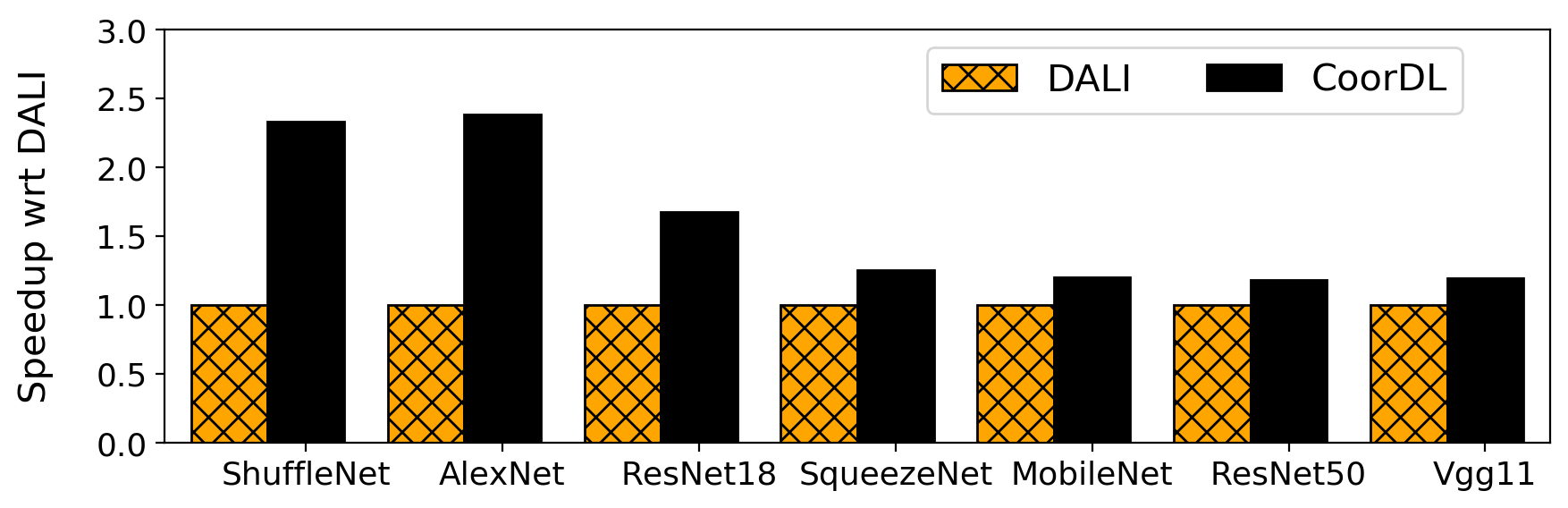}}
   
  \mycaption{HP search with ImageNet-22k dataset}{This plot shows the normalized training speed wrt DALI, when training 8 concurrent HP search jobs on \confssd. }
  \label{fig-eval-hp-img22k}
\end{figure}

\subsection{Evaluation with ImageNet22k}
ImageNet-22k is the extended version of the
popular ImageNet-1K dataset, and contains
about 14 million images that belong
to 21841 different categories~\cite{imagenet22k}. The average 
size of an image in this dataset is about 
90KB, much smaller than the average image size
in OpenImages dataset (300KB), as well
as ImageNet-1K (150KB).

When we train the image classification models with
ImageNet-22k on \confssd, \minio results in
20\% higher cache hits than DALI-shuffle that resulted in
1.5\myx faster training  on ShuffleNet, and
1.4\myx faster on AlexNet and ResNet18.

Next, when we perform distributed training on these
models on \confssd with 2 servers, AlexNet
trained 1.3\myx faster, Shufflenet trained 1.33\myx faster 
and ResNet18 achieved 1.12\myx speedup.
The fetch stalls with ImageNet-22k was
lower than a more complex dataset like
OpenImages because of the low per-image
size that increased the the number of samples
the storage can deliver per second. 

Finally, we perform HP search with eight 
concurrent jobs on \confssd on the seven
image classification models. As 
shown in Figure~\ref{fig-eval-hp-img22k},
\sysname results in upto 2.5\myx speedup.

\subsection{Cache misses with \sysname}
 \sysname's
\minio cache is designed to minimize the amount of
storage \vio per epoch, by efficiently utilizing the 
all the items in cache. Table~\ref{tbl-eval-fetch-miss}
enumerates the fetch misses and total disk \vio for DALI-seq,
DALI-shuffle and \sysname when training ShuffleNetv2 on OpenImages
dataset on \confssd. This server can cache 65\% of the dataset.
\sysname is able to reduce disk \vio by 47\% compared to DALI-seq and 
33\% compared to DALI-shuffle,
by reducing thrashing by 47\% and 33\% respectively.
\minio cache is able to reduce the cache misses down to capacity misses.


\setlength\mylength{\dimexpr.5\columnwidth-10\tabcolsep-0.5\arrayrulewidth\relax}
\begin{table}[!t]
  \small
  \centering
  \ra{1.3}
     \mycaption{Impact on fetch misses and disk IO}{When training
  	ResNet18 on OpenImages (645GB), \sysname reduces cache misses
  	from 66\% to 35\%. \confssd caches 65\% of the dataset, so this
  	is the minimum miss rate.}
  \vspace{-1em}
   \begin{tabular}{lrrr}
  	\toprule[1pt]
  	 & DALI-seq & DALI-shuffle & \sysname \\
  	\midrule
  
  	\emph{Cache miss}  & 66\% & 53\% & 35\%  \\
  	\emph{Disk IO (GB)} & 422 & 340 & 225\\

  	\bottomrule[1pt]
  \end{tabular}

\label{tbl-eval-fetch-miss}
\end{table}

\begin{figure}[bp!]
	\centering
  \begin{tabular}{l}
  \subfloat[Distributed training \label{fig-dist-mint-micro}]{\adjustbox{raise=-3.5pc}{\includegraphics[width=.28\textwidth]{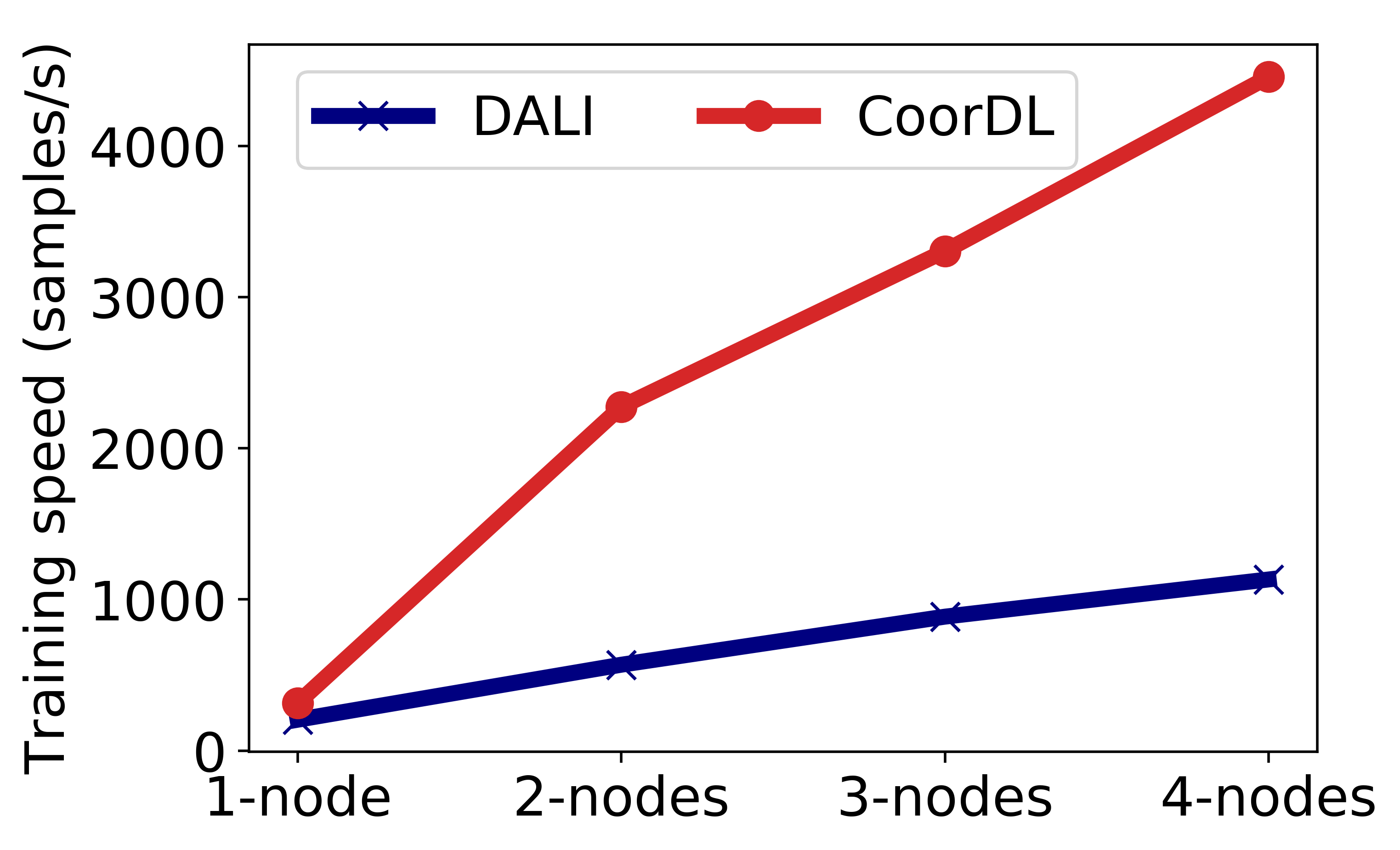}} }
  \quad
   \subfloat[Disk IO
  \label{tbl-dist-mint-io}]{
  		\begin{tabular}{ccc}
  			\toprule[1pt]
  			Nodes & Disk\\ 
  			& IO(GB)\\
  			\midrule
  			 1 & 342\\
 			 2 & 119\\
 			 3 & 70 \\
 			 4 & 50 \\
  			\bottomrule[1pt]
  		\end{tabular}
  }%

  \label{fig-eval-dist-mint-micro}
\end{tabular}
\mycaption{Distributed training with \sysname}{The plot compares DALI with \sysname when training ResNet50 across upto 4 nodes. Even when each node can cache 65\% of the dataset, DALI results in I/O bound training due to disk fetch, while \sysname results in zero disk accesses beyond first epoch.}
\end{figure}
\subsection{Scalability of partitioned caching}
Our paper shows that when training
is distributed across just enough servers that can cache the entire
dataset in memory, partitioned caching can
speed up training jobs by upto 15\myx. However, when we distribute training
to more servers, such that their aggregate memory is higher than
the total dataset size, \sysname continues to outperform
DALI as shown in Figure~\ref{fig-dist-mint-micro}. In this
experiment, we train ResNet50 on OpenImages on \confhdd,
where each server can cache 65\% of the dataset.
When training extends to 24 GPUs(3servers), or 32 GPUs(4 servers),
the throughput with \sysname increases because, training is
not bottlenecked on \vio and more GPUs for training naturally results
in faster training due to increase in GPU parallelism. With DALI, although 
the throughput increases, it is still \vio bound; the increase in throughput is due to the reduced disk \vio per server when training is distributed as shown in Table~\ref{tbl-dist-mint-io}. 
Although the \vio per server decreases with DALI as we distribute training 
across more servers, note that the GPU parallelism is also
proportionately increasing; the GPU compute rate ($G$) and prefetch rate ($F$),
are proportionately increasing, leaving the performance gap the same.

\begin{figure}[h!]

  {\includegraphics[width=0.42\textwidth]{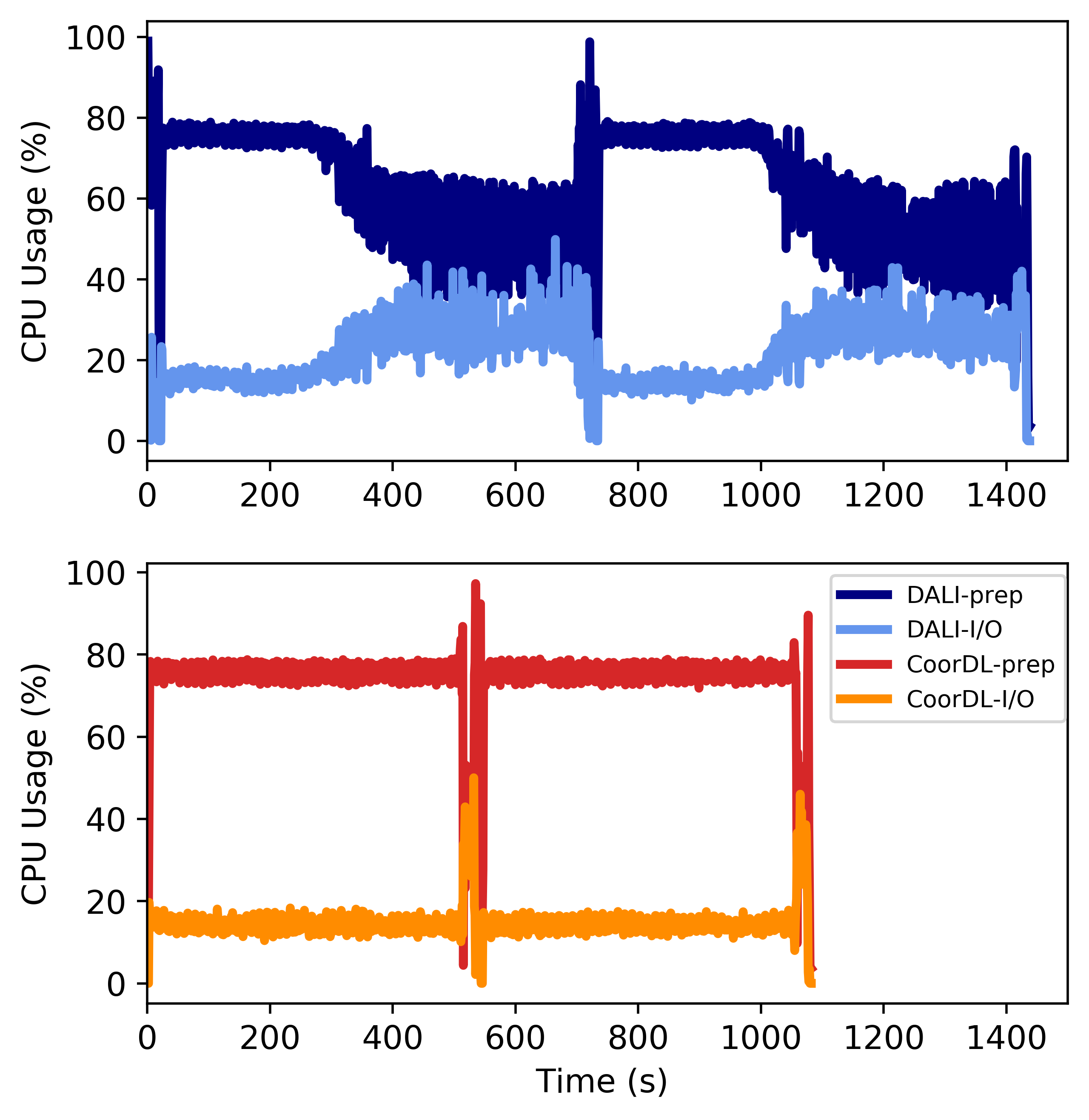}}
   \vspace{-1em}
  \caption{This plot shows the CPU utilization over time for DALI and \sysname when training ResNet18 on OpenImages. \sysname uses cache effectively to reduce disk I/O, therefore utilizing CPU on useful pre-processing rather than waiting on I/O}
  \label{fig-mint-cpu}

\end{figure}

\begin{figure}[h!]

  {\includegraphics[width=0.42\textwidth]{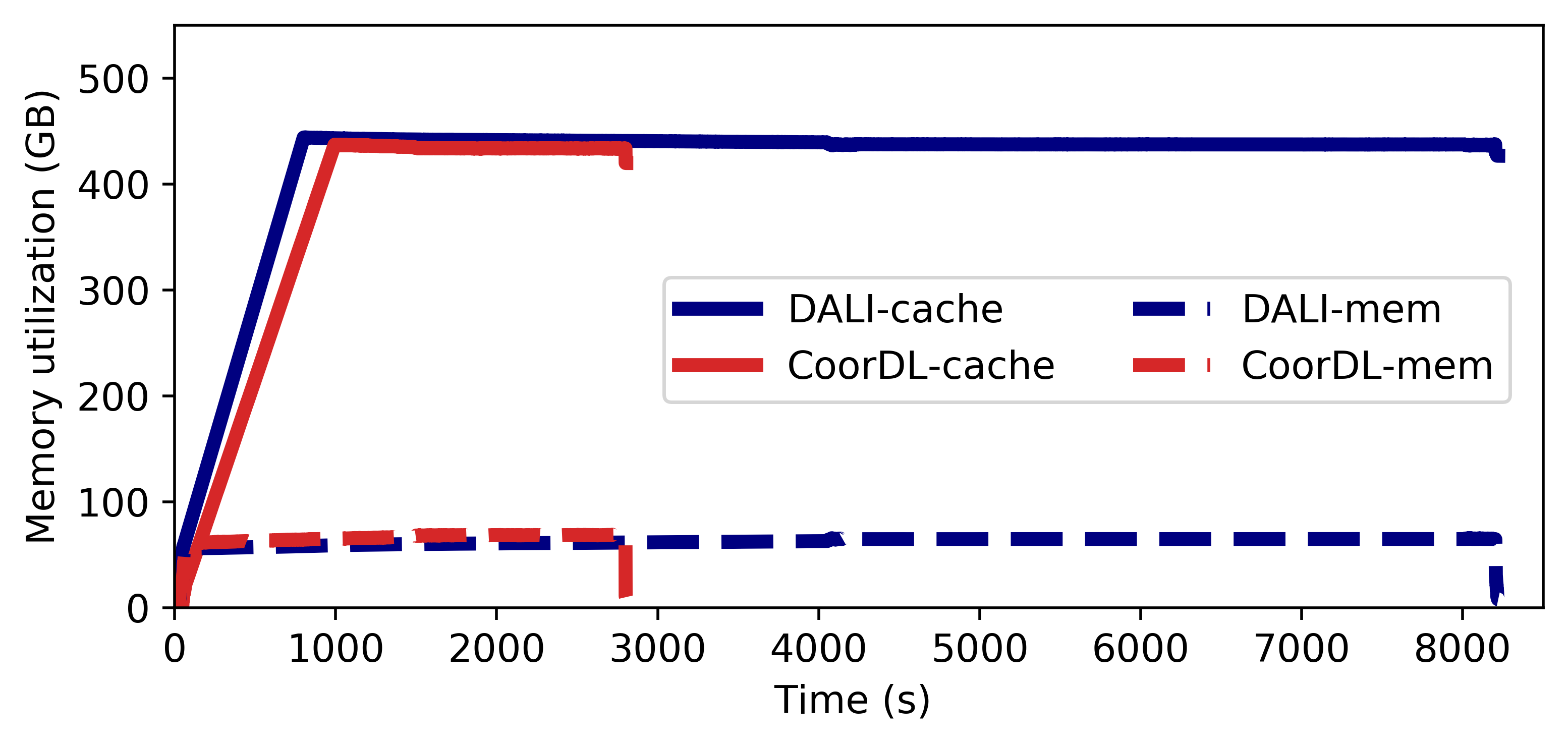}}
   \vspace{-1em}
  \caption{This plot shows the memory utilization for two epochs of HP search using AlexNet on OpenImages, with 8 concurrent jobs. \sysname uses 5GB of extra process memory; resulting in 5GB lower cache space.}
  \label{fig-eval-hp-mem}

\end{figure}

\subsection{HP search with fully cached dataset}
 The core of
\sysname's ability to speed up HP search jobs comes from coordinating
pre-processing to overcome the imbalance in the ratio of CPU cores to
GPU. We perform HP search with 8 jobs on \confssd with ImageNet-1k
dataset that fits entirely in memory. As shown in
Table~\ref{tbl-hp-cached},
\sysname sped up HP search by 1.9\myx on AlexNet and and 1.2\myx on ResNet50 by eliminating redundant pre-processing operations.

\setlength\mylength{\dimexpr.5\columnwidth-10\tabcolsep-0.5\arrayrulewidth\relax}
\begin{table}[!h]
  \small
  \centering
  \ra{1.3}
      \caption{On \confssd, when training with the small ImageNet-1k dataset that fits in memory, \sysname provides upto 1.87\myx speedup by eliminating redundant prep}
       \vspace{-1em}
   \begin{tabular}{lrrr}
  	\toprule[1pt]
  	& \multicolumn{2}{c}{Per job speed (Samples/s)}\\
  	 Model  & DALI & \sysname \\
  	\midrule
  	ShuffleNet & 1441 & 1.81\myx \\
	AlexNet & 1399 & 1.87\myx \\
	ResNet18 & 1056 & 1.53\myx \\
	SqueezeNet & 835 & 1.50\myx \\
	MobileNet & 752 & 1.35\myx \\
	ResNet50 & 569 & 1.21\myx \\
	VGG11 & 552 & 1.22\myx \\
  	\bottomrule[1pt]
  \end{tabular}

\label{tbl-hp-cached}
\end{table}

\subsection{Resource utilization with \sysname}
\vheading{CPU utilization with \sysname}.
The paper showed how \minio reduces the amount of
data fetched from storage in each epoch and
regularizes the data access pattern. 
Profiling the CPU during training of ResNet18 on OpenImages
shows that the pre-processing
threads in DALI are often stalled waiting for I/O as in Figure~\ref{fig-mint-cpu}. 
Since \minio
reduces the total disk I/O, \sysname is able to better utilize the CPU
threads for pre-processing. The combination of lower disk I/O and
better CPU utilization leads to shorter training times when using \sysname.

\vheading{Low memory overhead of co-ordinated prep}.  By design,
co-ordinated prep has the same memory requirements as DALI. To
experimentally validate this, we track the memory utilization of
running hyperparameter search on AlexNet on OpenImages on a \confssd
server using eight concurrent jobs.
Figure ~\ref{fig-eval-hp-mem}
plots the memory utilization over time for both the process working
memory, and the cache.
\sysname uses 5 GB of extra process memory to store prepared
mini-batches in memory until all hyperparameter jobs consume it. We
reduce the cache space given to \sysname by 5 GB (keeping the total
memory consumption same for \sysname and DALI). Despite the lower
cache space, \sysname still accelerated training by 2.9\myx.

\section{Building \pysysname in native PyTorch}
As a proof of concept, we implemented two 
of the techniques behind \sysname, \minio and
\emph{coordinated prep} as a pluggable
module to the native PyTorch DL (without
DALI). This section
briefly describes the implementation and
presents the evaluation of \pysysname against 
the native PyTorch DL.

 \begin{figure*}[!tbh]
  \vspace{-3mm}
  \centering

   \vspace{-3mm}
   \subfloat[HDD\label{fig-mcache-hdd}]{{\includegraphics[width=.45\linewidth]{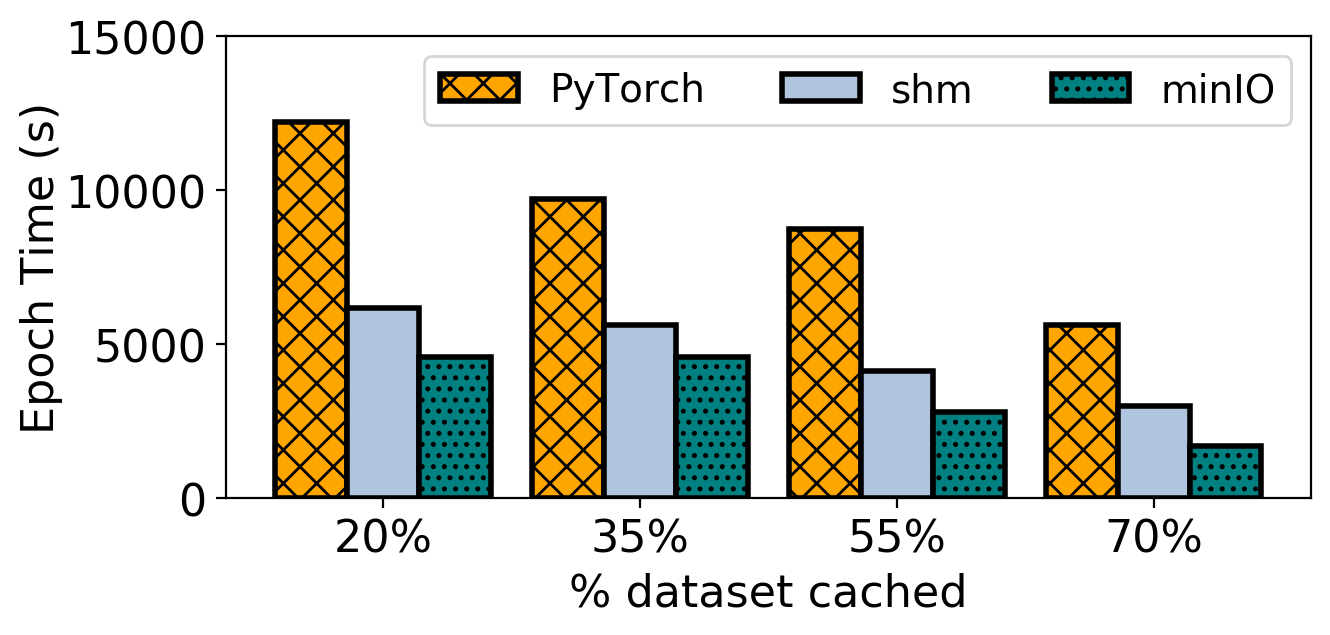} }}%
  \quad
   \vspace{-3mm}
  \subfloat[SSD\label{fig-mcache-ssd}]{{\includegraphics[width=.45\linewidth]{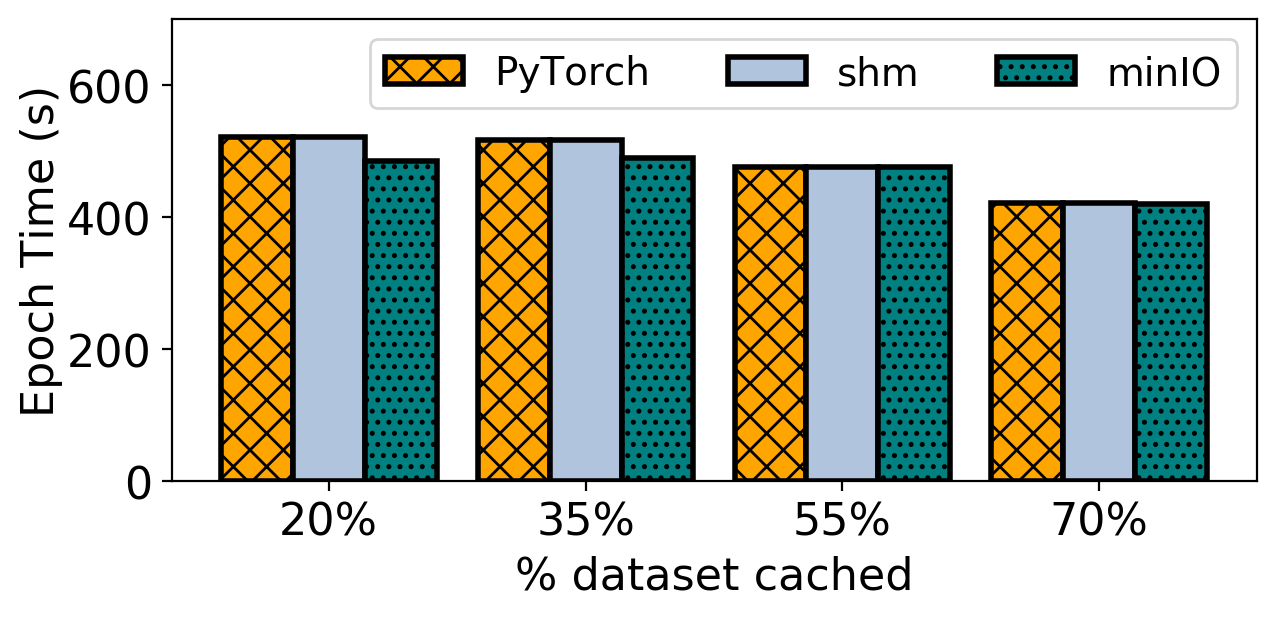} }}%
  
    \mycaption{Evaluation of \minio caching policy}{The graphs compare the native PyTorch DataLoader with \pysysname's \minio caching policy, and shows the speedup due to the two components in \minio; sequential access in shared memory(shm), and increased cache hits(\minio). On HDDs the sequential access of image files in shm provides significant speedup. On SSD, benefits with \minio are marginal because training is bottlenecked on CPU prep.}
  \label{fig-eval-mcache}

\end{figure*}
\subsection{Implementation}
 \pysysname is implemented as a pluggable \dl module for 
PyTorch with minimal changes to its current \dl API. 
\pysysname is implemented 
in 650 lines of Python code. 
\pysysname is implemented using Python's
shared memory abstraction because
PyTorch spawns multiple processes instead
of threads to parallelize data fetch and prep
(due to Python's Global Interpreter Lock
limiting concurrency of threads).

\subsection{Evaluation}
We evaluate \pysysname on a server with 8 \vtt{V100} GPUs, each with 16GB of GPU DRAM. Our server is  2 socket, 14-core Intel Xeon E5-2690@2.6GHz, with 500GB DRAM, and 2 different storage devices (SSD and HDD). We evaluate \pysysname on five image classification models; AlexNet~\cite{krizhevsky2012imagenet}, ResNet18~\cite{he2016deep}, ShuffleNetv2~\cite{zhang2018shufflenet}, SqueezeNet~\cite{iandola2016squeezenet} and MobileNetv2~\cite{sandler2018mobilenetv2}. We set the batch size to the maximum that fits the GPU (512 for Alexnet, Shufflenet and ResNet18, 256 for the others) . We train the
model for 5 epochs and report the average epoch time excluding the
first warmup epoch. We use the ImageNet 1K dataset of size 146\gb~\cite{imagenet1k} and PyTorch 1.1. To evaluate the benefits of \pysysname, we run our jobs in a Docker container with restricted memory to mimic the scenario where the dataset does not entirely fit in DRAM.  This is equivalent to running the full ImageNet dataset (22K classes - 1.3TB) on our server.

\subsubsection{\textbf{Multi-GPU training in a server}}
We now evaluate the benefit of using \pysysname to perform multi-GPU training in a single server.

\vheading{Hard drives}.
Figure~\ref{fig-mcache-hdd} plots the stabilized per epoch time as a function of cache size for ResNet18. In this experiment, DataParallel training is performed on 8 GPUs, each with a batch size of 512 and a total of 24 data workers pre-processing  in parallel. \pysysname brings down the per-epoch training time by 2.1\myx - 3.3\myx. This is due to two reasons. First, \pysysname increases the sequentiality of reading data items from the disk by indexing the entire data item instead of individual pages. Each data item in the ImageNet dataset is on average 150KB, which spans about 28 pages on disk. The native PyTorch DL fetches the pages of data items on demand, whenever the CPU thread requires to decode the item. As multiple data workers decode images in parallel, the pages from different images were requested randomly. \pysysname reduces this randomness by reading the entire data item into memory, before decoding it. 
Second, \minio caching policy results in 20\% lower cache misses as compared to the page cache's LRU scheme. Given the low throughput of disks (15MBps), this translates to a high savings in training time.

\vheading{Solid state drives}. Figure~\ref{fig-mcache-ssd} shows the
variation in training time for different cache sizes, when the dataset
is accessed from a fast solid state drive (SSD). The throughput of the
SSD is 500MBps. Reduction in cache trashing does not reduce the
training time significantly because we are bottlenecked on
pre-processing at the CPU (pre-processing throughput is around
327MBps).  Therefore, the 20\% reduction in store misses translates to
a mere 7\% lower training time. Note that when an
optimized library like DALI is used
for pre-processing, the CPU prep rate increases, making
storage the bottleneck; this makes \minio's savings more
significant with DALI.

 \begin{figure}[!t]
  \centering 
  \includegraphics[width=.45\textwidth]{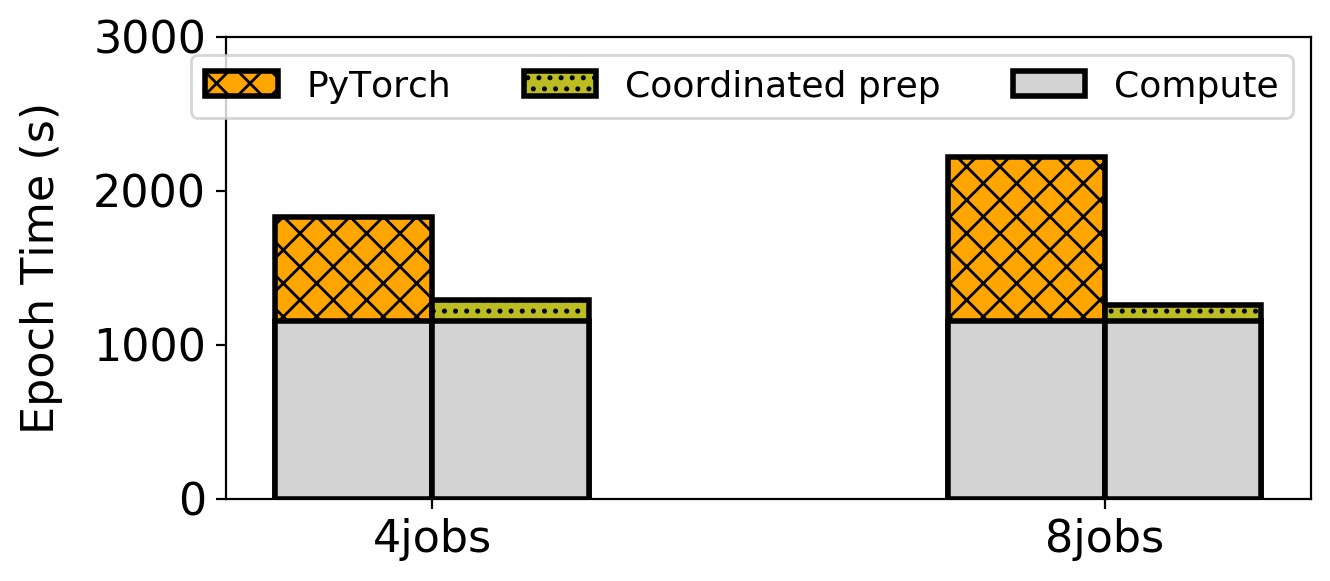}
   \mycaption{Evaluation of coordinated prep}{This graph compares PyTorch against \pysysname's  coordinated prep when training 6 or 8 HP search jobs on a server. Coordinated prep is able to reduce prep stalls significantly
   as compared to PyTorch}
    \label{fig-ucache-micro}

\end{figure}

\subsubsection{\textbf{HP Search}}

\label{sec-appendix-ucache}
To evaluate the benefits of coordinated prep, we construct a microbenchmark where each job trains the ResNet18 model on a single GPU in a server, when the entire dataset is cached in memory. We evaluate two scenarios; 4 jobs, each using 6 data workers for pre-processing ( 4 GPU and 24 CPU), and 8 jobs with 3 data workers each ( 8 GPU and 24 CPU). The per-epoch time for these scenarios is shown in Figure~\ref{fig-ucache-micro}. As the number of concurrent jobs increase, the data stall time increases because each job gets fewer CPU cores for pre-processing. \pysysname reduces the data stall time close to 0 in both cases.   It does so by launching a unified data loading process that pre-processes the dataset exactly once per epoch using all 24 CPUs, and shares the prepared batches across all the jobs. This technique results in 1.8\myx lower training time when 8 jobs are run concurrently on a single server.

 \begin{figure}[!tbh]
  \centering 
    \vspace{-5mm}
      \subfloat[End-to-end workload on HDD\label{fig-e2e-hdd}] {{\includegraphics[width=.5\textwidth]{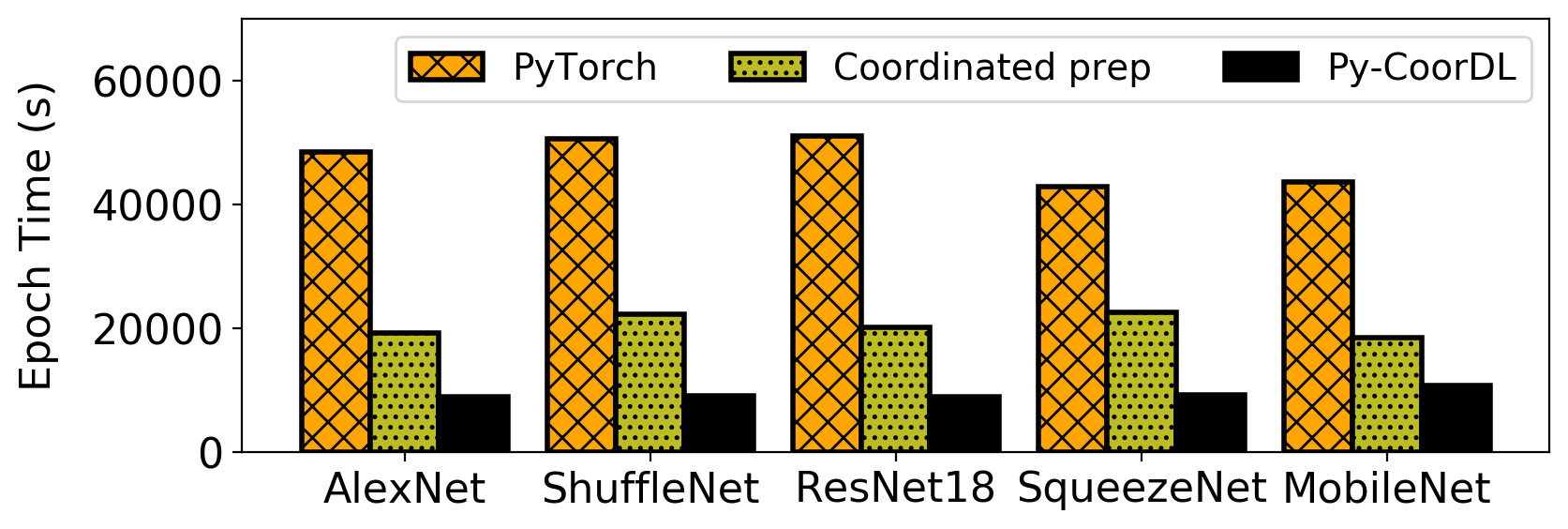} }}%
       \vspace{1em}
       \subfloat[End-to-end workload on SSD\label{fig-e2e-ssd}] {{\includegraphics[width=.5\textwidth]{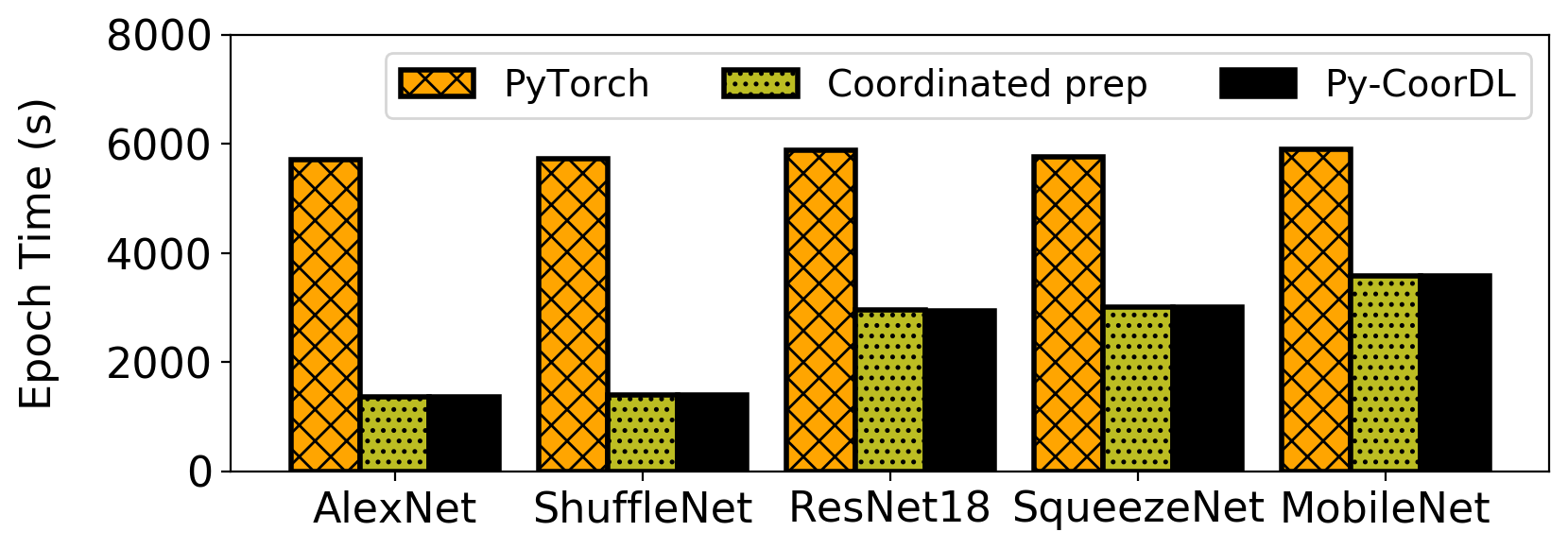} }}%
       \vspace{-3mm}       
         \mycaption{End-to-end evaluation}{The graphs compare the total search time for HP optimization on Ray Tune using the baseline PyTorch DL and \pysysname on hard disks and solid state drives. It also shows the 
         contribution of individual components; when just coordinated prep is used without \minio (indicated as coordinated prep) and when both techniques are used (shown as \pysysname). On SSDs, \minio does not help accelerate training as much as it does on HDDs, because the fetch rate is higher than the CPU prep rate on SSD.}
  \label{fig-eval-e2e}

\end{figure}
\subsubsection{\textbf{End-to-end benefit of \pysysname}}

\label{sec-appendix-e2e}
We now evaluate the end-to-end benefit of \sysname using a macrobenchmark; HP search using Ray Tune~\cite{liaw2018tune} when dataset does not fit entirely in memory.

\vheading{Ray Tune}. Ray Tune~\cite{liaw2018tune} is a HP optimization framework that provides the flexibility of using various search algorithms such as Population Based Training (PBT), Median Stopping Rule, and HyperBand. Ray Tune uses one of these algorithms to pick a unique value for the HP, and launches a training job on one of the available GPUs. We modified Ray Tune's job executor to use  \pysysname and launch training jobs one on each available GPU in a server. We used the Hyperband search algorithm to sample 16 values of (learning rate, momentum) pairs and set the stopping criteria to be the completion of one epoch for brevity. The trends remain the same if the stopping criteria is set to a target accuracy. 

\vheading{Experiment setting}. We run this macrobenchmark on a machine with 8 GPUs (8 samples are trained in parallel). For the PyTorch DL, we set the number of data workers to 3 for each job and to evaluate \pysysname, we set data workers to 24. Note the total number of data workers in the system is the same in both cases. We set the cache size set to 110GB ($\approx$ 75\% of the dataset). We record the total reduction in search time compared to the baseline, and the contribution of each of our techniques, \emph{coordinated prep} alone, and when coordinated prep is combined with \minio caching. We evaluate the benefits of \pysysname in two scenarios; when the dataset resides on a slower storage media like hard drive and when it is on a relatively faster media like solid state drive. 

\vheading{Dataset resides on hard drive}. As shown in Figure~\ref{fig-e2e-hdd}, coordinated prep alone results in upto 2.5\myx speedup in total search time by reducing the total disk accesses by 2.5\myx. The savings in time comes directly from the reduced disk accesses because the \dl in this case is bottlenecked on \vio rather than pre-processing. When \minio caching policy is enabled, the effective speedup is close to 5.5\myx due to reduced storage miss and reduction in random accesses. 

\vheading{Dataset resides on solid state drive}. When the dataset is on a faster medium like SSD, whose throughput is higher than that of pre-processing, the bottleneck in the \dl shifts to CPU. In this scenario, as shown in Figure~\ref{fig-e2e-ssd}, coordinated prep reduces the overhead of pre-processing and speeds up search time by reusing prepared minibatches across jobs. With the addition of \minio policy, the search does not speed up significantly due to cheap IO.

\subsection{Summary} \pysysname speeds up DNN training jobs by 2\myx - 5.7\myx by enabling efficient reuse of both raw data items and pre-processed batches. Although 
\pysysname has marginal gains when 
dataset resides on SSD, the reason was the slow
pre-processing rate of data transformation
operations used by PyTorch DL. If prep rate goes up, fetch stalls become 
prominent, and \minio comes to the rescue, which is the case when using DALI for pre-processing~\cite{noauthor_fast_2019}.

	}{}
	
\end{document}